\documentclass[11pt]{article}
\usepackage{amssymb}

\usepackage{latexsym}
\usepackage{amsthm}
\usepackage{amsmath}
\usepackage[utf8]{inputenc}
\usepackage{fullpage}
\usepackage{boxedminipage}
\usepackage{listings}
\usepackage{minitoc}
\usepackage[pdftex]{graphicx}
\usepackage{graphicx}
\usepackage{nccmath}
\usepackage{amscd,graphicx,psfrag}
\usepackage{epic, xypic}
\usepackage{booktabs}

\numberwithin{equation}{section}

\newcommand{\be}{\begin{equation}}
\newcommand{\ee}{\end{equation}}
\newcommand{\bes}{\begin{equation*}}
\newcommand{\ees}{\end{equation*}}
\newcommand{\eqn}{\begin{eqnarray}}
\newcommand{\feqn}{\end{eqnarray}}
\newcommand{\eqnn}{\begin{eqnarray*}}
\newcommand{\feqnn}{\end{eqnarray*}}

\makeatletter \@addtoreset{equation}{section} \makeatother

\newif\ifpdf \ifx\pdfoutput\undefined \pdffalse
\else \pdfoutput=1 \pdftrue \fi \ifpdf \else \fi
\input epsf

\begin{document}

\ifpdf\DeclareGraphicsExtensions{.pdf, .jpg, .tif} \else%
\DeclareGraphicsExtensions{.eps, .jpg} \fi
\begin{titlepage}

    \thispagestyle{empty}
    \begin{flushright}
        \hfill{CERN-PH-TH/2012-020}\\
    \end{flushright}

    \vspace{30pt}
    \begin{center}
        { \huge{\textbf{Magic Coset Decompositions}}}

        \vspace{30pt}

        {\large{\bf Sergio L. Cacciatori$^{1,4}$, Bianca L. Cerchiai$^{2,4}$, and \ Alessio Marrani$^3$}}

        \vspace{80pt}

        {$1$ \it Dipartimento di Scienze ed Alta Tecnologia,\\Universit\`a degli Studi dell'Insubria,
Via Valleggio 11, 22100 Como, Italy\\
\texttt{sergio.cacciatori@uninsubria.it}}

        \vspace{10pt}

        {$2$ \it Dipartimento di Matematica,\\
Universit\`a degli Studi di Milano,  Via Saldini 50, 20133 Milano,
Italy\\
\texttt{bianca.cerchiai@unimi.it}}

        \vspace{10pt}

        {$3$ \it Physics Department,Theory Unit, CERN, \\
        CH 1211, Geneva 23, Switzerland\\
        \texttt{alessio.marrani@cern.ch}}

\vspace{10pt}

        {$4$ \it INFN, Sezione di Milano\\
Via Celoria, 16, 20133 Milano,
Italy}

        \vspace{85pt}
\end{center}

\vspace{5pt}

\begin{abstract}
By exploiting a ``mixed" non-symmetric Freudenthal-Rozenfeld-Tits
magic square, two types of coset decompositions are analyzed for the non-compact special K%
\"{a}hler symmetric rank-$3$ coset $E_{7\left( -25\right) }/\left[
\left( E_{6\left( -78\right) }\times U\left( 1\right) \right)
/\mathbb{Z}_{3}\right]
$ , occurring in supergravity as the vector multiplets' scalar manifold in $%
\mathcal{N}=2$, $D=4$ \textit{exceptional} Maxwell-Einstein theory.

The first decomposition exhibits maximal manifest covariance,
whereas the
second (\textit{triality-symmetric}) one is of Iwasawa type, with maximal $%
SO\left( 8\right) $ covariance.

Generalizations to \textit{conformal non-compact}, real forms of
non-degenerate, simple groups ``of type $E_{7}$'' are presented for
both classes of coset parametrizations, and relations to rank-$3$
simple Euclidean Jordan algebras and normed trialities over division
algebras are also discussed.
\end{abstract}

\end{titlepage}
\newpage \tableofcontents 

\section{\label{Introduction}Introduction}

The role of groups in Physics is at least threefold. First, they represent
symmetries that, by definition, introduce elegance in all the equations
which are manifestly symmetry invariant. If that was all, one may argue
that this would be a poor advantage. But symmetries also arise as
fundamental principles in constructing new theories, like, for example,
gauge symmetries for the Standard Model of particle physics, conformal
symmetry for string theory, or general covariance for the Einstein theory of
relativity.
Finally, symmetries, and then groups, play a key role in solving the
equations of motion.

A particular class is represented by the (semi)simple Lie groups (and
corresponding Lie algebras), which, once more, find application in a large number of mathematical and physical fields. All the finite dimensional complex Lie algebras have been classified by Wilhelm Killing, whose proofs have been made rigorous by \'{E}lie Cartan, who has also extended the classification to the non-compact, real cases. The well known result is that this classification has led to the discovery, beyond the famous classical series, of five exceptional algebras (of course together with the corresponding real forms):
$\frak{g}_{2}$, $\frak{f}_{4}$, $\frak{e}_{6}$, $\frak{e}_{7}$ and $\frak{e}%
_{8}$.

Despite their sporadicity, the appearance of exceptional Lie groups (and
algebras) in physics is anything but sporadic \cite{ramond}. The importance of compact
exceptional Lie groups in realizing grand unification gauge theories and
consistent string theories is well recognized. Similarly, the relevance of
non-compact real forms for the study of locally supersymmetric theories of gravity is well known. Other examples include sigma models based on quotients of exceptional Lie groups, which are of interest for string theory and conformal field theory applications as well. It is worth mentioning that the analysis of quantum criticality in Ising chains and the
structure of magnetic materials such as Cobalt Niobate has also
recently (and strikingly) turned out to be related to exceptional Lie groups of type $E$ (see \textit{e.g.}  \cite{science} and \cite{cobalt}, respectively).\medskip

Several properties of exceptional groups and algebras can be already
inferred from abstract theoretical considerations. Nevertheless, it is often
important to have explicit concrete realizations of the groups available in
term of matrices, for both numerical or analytical calculations. For
example, one could test conjectures related to confinement in non-Abelian
gauge theories (see \textit{e.g.} \cite{confinement}), and, more in general, perform explicit non-perturbative
computations in exceptional lattice GUT theories and in random matrix
theories.

In particular, among all the exceptional groups, there are specific motivations for physics to be interested in $E_{7}$ : recently, a strict relation between cryptography and black hole physics based on $E_{7}$ (and $E_{6}$) exceptional supergravity has been discovered
\cite{Duff-QIT, 4-qubits, ICL-4, Cerchiai-VG}.
However, actual computation of entangled expectation values requires again an explicit determination of the Haar measure and of the range of the parameters.
Moreover, fascinating group theoretical structures arise clearly in the description of the Attractor Mechanism for black holes in the Maxwell-Einstein supergravity \cite{AM-Refs}, such as the so-called magic exceptional supergravity \cite{GST} we are focusing on in the present investigation, which is related to the \textit{minimally non-compact} real $E_{7\left( -25\right) }$ form \cite{dobrev} of $E_{7}$.

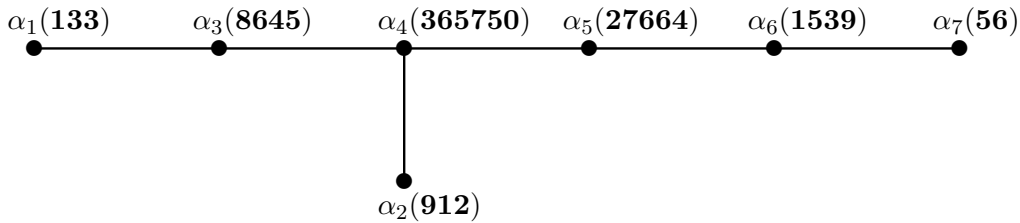
\begin{figure}[h]
\label{fig}\centering
\begin{picture}(200, 90)(-50,-70)%
\put(-160,10){\makebox(0,0)[l]{{$\alpha_1$({\bf 133})}}}
\put(-90,10){\makebox(0,0)[l]{{$\alpha_3$({\bf 8645})}}}
\put(-20,10){\makebox(0,0)[l]{{$\alpha_4$({\bf 365750})}}}
\put(50,10){\makebox(0,0)[l]{{$\alpha_5$({\bf 27664})}}}
\put(120,10){\makebox(0,0)[l]{{$\alpha_6$({\bf 1539})}}}
\put(190,10){\makebox(0,0)[l]{{$\alpha_7$({\bf 56})}}}
\put(-20,-60){\makebox(0,0)[l]{{$\alpha_2$({\bf 912})}}}
\put(-150,0){\circle*{6}}
\put(-80,0){\circle*{6}}
\put(-10,0){\circle*{6}}
\put(60,0){\circle*{6}}
\put(130,0){\circle*{6}}
\put(200,0){\circle*{6}}
\put(-10,-50){\circle*{6}}
\thicklines
\drawline(-150,0)(200,0)
\drawline(-10,0)(-10,-50)
\end{picture}
\caption{Dynkin diagram for $\frak{e}_{7}$}
\end{figure}

Before proceeding further, it is worth recalling some basic facts on the Lie algebra $\frak{e}_7$ of $E_{7}$. Let us start by stating that $\frak{e}_{7}$ is the unique exceptional Lie algebra of rank $7$, and it is characterized by the Dynkin diagram drawn in Fig. \ref{fig},
in which each dot corresponds to a simple root $\alpha _{i}$. These are free
generators of the root lattice $\Lambda _{R}=\sum_{i}\mathbb{Z}\alpha _{i}$.
The space $H^{\ast }=\Lambda _{R}\otimes \mathbb{R}$ is endowed with a
positive definite inner product $(\cdot|\cdot)$. The weight lattice $\Lambda _{W}$ is
the dual of $\Lambda _{R}$ with respect to the hooked product, which means
that it is freely generated over $\mathbb{Z}$ by the fundamental weights $%
\lambda ^{i}\in H^{\ast }$, $i=1,\ldots ,7$ defined by $\langle \alpha
_{i},\lambda ^{j}\rangle =\delta _{i}^{j}$, with:
\begin{equation}
\langle \alpha ,\lambda \rangle :=2\frac{(\alpha |\lambda )}{(\alpha |\alpha
)}.
\end{equation}
There is a univocal correspondence between fundamental weights and  fundamental representations, and all the irreducible finite dimensional
representations can be generated from the basic ones, which are
indicated in parenthesis in Fig. \ref{fig}. Here, we are going to deal with
the two lower dimensional, namely the fundamental $\mathbf{56}$
and the adjoint $\mathbf{133}$.\newline

The complex algebra $\frak{e}_{7}$ is completely characterized by its Dynkin
diagram, from which one can reconstruct the adjoint representation, that,
being faithful, is isomorphic to the algebra itself. Since $\frak{e}_{7}$ is
a $133$-dimensional complex algebra, it follows that such a representation
is the aforementioned $\mathbf{133}$.

The Lie algebra $\frak{e}_{7}$ exhibits four distinct non-compact, real forms. This means
that there are four inequivalent ways to select a $133$-dimensional real
subspace of the $266$-dimensional real space underlying the complex algebra $%
\frak{e}_{7}$, in such a way that the selected subspace endowed with the
inherited Lie product is itself a (real) Lie algebra. For each simple Lie algebra $%
\frak{g}$ there is a unique simply connected Lie group $G$ (up to
isomorphisms), such that $\frak{g}$ is the corresponding Lie algebra. The
complex Lie group $E_{7\left( \mathbb{C}\right) }$ contains a maximal
compact subgroup, which is a $133$-dimensional real Lie group (denoted as $%
E_{7\left( -133\right) }$), whose Lie algebra is then called the compact form
(denoted\footnote{%
The Killing form $K$ on a complex Lie algebra is defined by $K(X,Y):=\mathrm{%
Tr}(ad(X)ad(Y))$ and is non-degenerate for a simple algebra and on the
corresponding real forms. In particular, for a non-compact form it is
negative definite on the maximal compact subalgebra, namely on the maximal
Lie subalgebra, whose exponentiation generates a compact Lie (sub)group.} as $%
\frak{e}_{7(-133)}$), where in parenthesis the signature of the Killing form
(number of the positive eigenvalues minus number of the negative ones) is
indicated.

The non-compact, real forms are in correspondence with the maximal compact
subalgebras of $\frak{e}_{7(-133)}$ (\textit{i.e.}, the compact Lie
subalgebras that are not properly contained in a proper subalgebra of $%
\frak{e}_{7(-133)}$ itself). The same holds at group level. There are four
such subalgebras and therefore four corresponding real forms, which we collect in Table \ref{tab} (at Lie group level). For a recent treatment of $E_{7}$ groups (and
algebras), see \textit{e.g.} \cite{E7magic}.

\begin{table}[tbph]
\begin{center}
\begin{tabular}{|c|c|c|}
\hline &&\\[-1.7ex]
Symbol & Real Form & Maximal compact subgroup (mcs) \\[0.5ex]
\hline &&\\[-1.7ex]
$E_{7(-133)}$ & Compact & $E_{7(-133)}$ \\[0.5ex]
\hline&&\\[-1.7ex]
$E_{7(7)}$ & Split & $SU(8)/\mathbb{Z}_{2}$ \\[0.5ex]
\hline&&\\[-1.7ex]
$E_{7(-5)}$ & EVI & $(Spin(12)\times USp(2))/\mathbb{Z}_{2}$ \\[0.5ex]
\hline&&\\[-1.7ex]
$E_{7(-25)}$ & EVII & $(E_{6(-78)}\times U(1))/\mathbb{Z}_{3}$ \\[0.5ex]
\hline
\end{tabular}
\end{center}
\caption{The real forms of $E_{7}$.}
\label{tab}
\end{table}

The plan of the paper is as follows.

As anticipated, we are going to deal with the \textit{minimally non-compact} real
form of $\frak{e}_{7}$ ($E_{7}$), namely with $\frak{e}_{7\left( -25\right)
} $ and its corresponding Lie group $E_{7\left( -25\right) }$, both denoted
by EVII (see Table \ref{tab}). In Sec.~\ref{sec:56}, by starting from its general
construction through the Tits magic square, we study the Lie algebra $\frak{e}%
_{7\left( -25\right) }$ itself, and we explicily construct a
realization in the fundamental $\mathbf{56}$ representation embedded in $%
\frak{usp}(28,28)$. The matrix elements obtained with this technique
turn out to be strictly related to the invariant totally symmetric rank-$3$ so-called $d$-tensor
of the $E_{6(-78)}$ group, thus allowing for different expressions, depending on the choice of the basis for the relevant rank-$3$ (simple) Euclidean Jordan algebra.\newline

In the present paper, we focus on two remarkable explicit
parametrizations of the symmetric manifold\footnote{%
For previous studies of exceptional cosets in supergravity, see \textit{e.g.}
\cite{truini-bidenharn}.}
\begin{equation}
\mathcal{M}:=\frac{E_{7\left( -25\right) }}{K}=\frac{E_{7(-25)}}{\left(
E_{6\left( -78\right) }\times U\left( 1\right) \right) /\mathbb{Z}_{3}}
\label{M-call}
\end{equation}
(obtained by suitably exponentiating the corresponding coset Lie
algebra), which can be regarded as the classical vector multiplets'
scalar manifold of the $\mathcal{N}=2$, $D=4$ Maxwell-Einstein
so-called exceptional magic
supergravity theory, based on the rank-$3$ Euclidean simple Jordan algebra $%
\frak{J}_{3}\left( \mathbb{O}\right) $ on the normed division algebra of the
octonions $\mathbb{O}$ \cite{GST}.

The first type of coset parametrization/decomposition, analyzed in Sec. \ref
{manifest}, exhibits maximal manifest covariance with respect to the maximal
compact subgroup (\textit{mcs}) $E_{6\left( -78\right) }\times U\left(
1\right) $ of $E_{7(-25)}$ (up to $\mathbb{Z}_{3}$; see (\ref{M-call})).
Interestingly, such a coset parametrization, also exhibiting a manifest
complex (actually, special K\"{a}hler) structure, can be generalized to
encompass a more general class of Lie groups, which in Sec.~\ref
{Generalizations} we identify \textit{at least} as the \textit{conformal
non-compact} real forms of simple, non-degenerate Lie groups ``of type $%
E_{7} $'' \cite{Brown}, of which $E_{7\left( -25\right) }$ (in its $\mathbf{56}$
representation) can be considered as the generic representative. Groups ``of
type $E_{7}$'' have recently appeared in Theoretical Physics, in the
investigation of single - \cite{Duff-FD-1} and multi-centered \cite
{FMOSY-1,Irred-1,Small-1,FMY-FD-1,FMY-T-CV} extremal black hole solutions in
supergravity theories, as well as in the study of matter creation in the
Universe \cite{FK-creation}.

The second coset parametrization, studied in Sec. \ref{Iwa-Exc}, relies on
the Iwasawa construction, already analyzed for the split form $E_{7\left(
7\right) }$ \textit{e.g.} in \cite{Iwa-N=8}. In this case, the maximal
manifest covariance reduces down to an $SO(8)$ subgroup of $E_{7\left(
7\right) }$, which will interestingly turn out to be related to the
automorphism group Aut$\left( \mathbf{t}\left( \mathbb{O}\right) \right) $
of the \textit{normed triality} $\mathbf{t}\left( \mathbb{O}\right) $ over
the octonions $\mathbb{O}$ (entering the Tits' construction). The well known $%
SO(8)$ triality is manifest in such an approach, as detailed in the
group theoretical analysis of Subsecs. \ref{Manifest-Covariance} and \ref
{Group-Theory}. As discussed in Sec. \ref{Generalizations}, also this construction of the Iwasawa decomposition can be generalized \textit{at least} to the \textit{conformal non-compact} real forms of simple, non-degenerate Lie groups ``of type $E_{7}$''; the resulting manifest covariance is then given by an $SO\left( q\right) \times \mathcal{A}_{q}$ subgroup, which remarkably shares the same Lie algebra as the automorphism group Aut$\left( \mathbf{t}\left( \mathbb{A}\right) \right) $ of the \textit{normed triality} over the relevant normed division algebra (see \textit{e.g.} \cite{Hurwitz}) $\mathbb{A}=\mathbb{R}$ (reals), $\mathbb{C}$ (complex
numbers), $\mathbb{H}$ (quaternions), $\mathbb{O}$ (octonions).

Final remarks, comments and discussion of further possible developments are
given in the concluding Sec.~\ref{Conclusion}.


\section{\label{sec:56}The Lie algebra $\frak{e}_{7(-25)}$ in the $\mathbf{56%
}$}

In order to construct the Lie algebra $\frak{e}_{7(-25)}$, we are going to
follow a procedure similar to the one outlined in Sec. 7 of \cite{IY}, based
on the non-symmetric ``mixed'' \textit{magic square} \cite{bart-sud,GST,tits} displayed in Table \ref{square}
:

\begin{table}[h]
\begin{center}
\begin{tabular}{|c|c|c|c|c|}
\hline&&&&\\[-1.7ex]
& $\mathbb{R}$ & $\mathbb{C}$ & $\mathbb{H}$ & $\mathbb{O}$ \\[0.5ex] \hline&&&&\\[-1.7ex]
$\mathbb{R}$ & $SO(3)$ & $SU(3)$ & $USp(6)$ & $F_{4(-52)}$ \\[0.5ex] \hline&&&&\\[-1.7ex]
$\mathbb{C}$ & $SU(3)$ & $SU(3)\times SU(3)$ & $SU(6)$ & $E_{6(-78)}$ \\[0.5ex]
\hline&&&&\\[-1.7ex]
$\mathbb{H}_{S}$ & $Sp(6,\mathbb{R})$ & $SU(3,3)$ & $SO^{\ast }(12)$ & $%
E_{7(-25)}$ \\[0.5ex] \hline&&&&\\[-1.7ex]
$\mathbb{O}_{S}$ & $F_{4(4)}$ & $E_{6(2)}$ & $E_{7(-5)}$ & $E_{8(-24)}$ \\[0.5ex]
\hline
\end{tabular}
\end{center}
\caption{The ``mixed" magic square.}
\label{square}
\end{table}

The rows and the columns contain the division algebras of the real numbers $%
\mathbb{R}$, the complex numbers $\mathbb{C}$, the quaternions $\mathbb{H}$
and the octonions $\mathbb{O}$. Since at the group level we focus on the
(minimally) \textit{non-compact} form $E_{7(-25)}$, we need to start from
the split form $\mathbb{H}_{S}$ of the quaternions in the third row. On the
other hand, we are also interested in identifying explicitly its maximal
compact subgroup $K:=E_{6\left( -78\right) }\times U(1)/\mathbb{Z}_{3}$ \cite
{IY}, and therefore the usual form $\mathbb{C}$ of the complex field in the
second row is to be considered.

The Tits' formula then yields the Lie algebra $\mathcal{L}$ corresponding to
division algebras in row $\mathbb{A}$ and column $\mathbb{B}$ as follows \cite{tits}:
\begin{equation}
\mathcal{L}\left( \mathbb{A},\mathbb{B}\right) =\text{Der}\left( \mathbb{A} \right) \oplus \text{Der}%
\left( \frak{J}_{3}\left( \mathbb{B}\right) \right) \dotplus \left( \mathbb{A}^{\prime
}\otimes \frak{J}_{3}^{\prime }\left( \mathbb{B}\right) \right) .
\label{Tits-formula}
\end{equation}
The symbol $\oplus $ denotes direct sum of algebras, whereas $\dotplus $
stands for direct sum of vector spaces. Moreover, Der are the linear
derivations, $\frak{J}_{3}\left( \mathbb{B}\right) $ denotes the rank-$3$ Jordan
algebra on $\mathbb{B}$, and the priming amounts to considering only traceless
elements.

In particular, for the Lie algebra of $E_{7(-25)}$ the Tits' formula (\ref
{Tits-formula}) reads:
\begin{equation}
\frak{e}_{7\left( -25\right) }=\mathcal{L}\left( \mathbb{H}_{S},\mathbb{O}%
\right) =\text{Der}(\mathbb{H}_{S})\oplus \text{Der}(\frak{J}_{3}(\mathbb{O}%
))\dotplus (\mathbb{H}_{S}^{\prime }\otimes \frak{J}_{3}^{\prime }(\mathbb{O}%
)).  \label{algebra1}
\end{equation}
$\mathbb{H}_{S}^{\prime }$ denotes the \textit{imaginary} split quaternions,
and the following multiplication rule holds for the units $i,j,k\in \mathbb{H%
}_{S}$ (\textit{cfr.} \textit{e.g.} (A.18) of \cite{G-1}):
\begin{equation}
i\,j=k=-j\,i,\quad j\,k=-i=-k\,j,\quad k\,i=j=-i\,k,\quad i^{2}=-1,\quad
j^{2}=k^{2}=1.
\end{equation}
An inner product can be defined on $\mathbb{H}_{S}$ as:
\begin{equation}
\langle h_{1},h_{2}\rangle :=\text{Re}(\bar{h}_{1}h_{2}),~h_{1},h_{2}\in
\mathbb{H}_{S}.
\end{equation}
Also, notice that if $L$ and $R$ respectively are the left and right
translation in $\mathbb{H}_{S}$, then a derivation $D_{h_{1},h_{2}}\in $Der$(%
\mathbb{H}_{S})$ can be constructed from $h_{1},h_{2}\in \mathbb{H}_{S}$ as:
\begin{equation}
D_{h_{1},h_{2}}:=[L_{h_{1}},L_{h_{2}}]+[R_{h_{1}},R_{h_{2}}].
\end{equation}

The rank-$3$ octonionic Jordan algebra $\frak{J}_{3}(\mathbb{O})$ is defined
as the algebra of the $3\times 3$ hermitian matrices of the form:
\begin{equation}
J=\left(
\begin{array}{ccc}
a_{1} & o_{1} & o_{2} \\
o_{1}^{\ast } & a_{2} & o_{3} \\
o_{2}^{\ast } & o_{3}^{\ast } & a_{3}
\end{array}
\right)
\end{equation}
with $a_{i}\in \mathbb{R}$, and $o_{i}\in \mathbb{O}$, $i=1,2,3$. The Jordan
product $\circ $ is thus realized as the symmetrized matrix multiplication:
\begin{equation}
j_{1}\circ j_{2}:=\frac{1}{2}(j_{1}j_{2}+j_{2}j_{1}),~j_{1},j_{2}\in \frak{J}%
_{3}(\mathbb{O}).
\end{equation}
It is then possible to introduce an inner product on the Jordan algebra:
\begin{equation}
\langle j_{1},j_{2}\rangle :=\text{Tr}(j_{1}\circ j_{2}).
\end{equation}
Furthermore, there is a cubic form, which is defined for any $%
j_{1},j_{2},j_{3}\in \frak{J}_{3}(\mathbb{O})$ as \cite{freudenthal} (for a
recent account, see \textit{e.g.} \cite{Small-Orbits-Phys,Small-Orbits-Maths}):
\begin{eqnarray}
Det(j_{1},j_{2},j_{3}) &:&=\frac{1}{3}\text{Tr}(j_{1}\circ j_{2}\circ j_{3})-%
\frac{1}{6}\left( \text{Tr}(j_{1})\text{Tr}(j_{2}\circ j_{3})+\text{Tr}%
(j_{2})\text{Tr}(j_{1}\circ j_{3})+\text{Tr}(j_{3})\text{Tr}(j_{1}\circ
j_{2})\right)  \notag \\
&&+\frac{1}{6}\text{Tr}(j_{1})\text{Tr}(j_{2})\text{Tr}(j_{3}).
\label{determinant}
\end{eqnarray}
In turn, this induces an action $\rhd $ of $\frak{J}_{3}(\mathbb{O})$ on
itself through $Det(j_{1},j_{2},j_{3}):=\frac{1}{3}$Tr$\left( (j_{1}\rhd
j_{2})\circ j_{3}\right) $, which by definition (\ref{determinant}) reads:
\begin{equation}
j_{1}\rhd j_{2}:=j_{1}\circ j_{2}-\frac{1}{2}\text{Tr}(j_{1})j_{2}-\frac{1}{2%
}\text{Tr}(j_{2})j_{1}+\frac{1}{2}\text{Tr}(j_{1})\text{Tr}(j_{2})I_{3}-%
\frac{1}{2}\text{Tr}(j_{1}\circ j_{2})I_{3},  \label{jaction}
\end{equation}
with $I_{3}$ the $3\times 3$ identity matrix.

An important ingredient entering Eq. (\ref{Tits-formula}) is the \textit{Lie
product} $[\cdot,\cdot]$, which in the case under consideration extends the
multiplication structure also to $\mathbb{H}_{S}^{\prime }\otimes \frak{J}%
_{3}^{\prime }(\mathbb{O})$; its general explicit expression can be found
\textit{e.g.} in Eq. (2.5) of \cite{E7magic}:
\begin{equation}
\lbrack h_{1}\otimes j_{1},h_{2}\otimes j_{2}]:=\frac{1}{12}\langle
j_{1},j_{2}\rangle D_{h_{1},h_{2}}-\langle h_{1},h_{2}\rangle \lbrack
L_{j_{1}},L_{j_{2}}]+\frac{1}{2}[h_{1},h_{2}]\otimes (j_{1}\circ j_{2}-\frac{%
1}{3}\langle j_{1},j_{2}\rangle I_{3}).
\end{equation}
It is known (see \textit{e.g.} \cite{G-1}, \cite{G-D=3})
that:
\begin{eqnarray}
\text{Der}\left( \frak{J}_{3}(\mathbb{O})\right) &\sim &\frak{f}_{4\left(
-52\right) }; \\
\text{Der}\left( \mathbb{H}_{S}\right) &\sim &\frak{sl}\left( 2,\mathbb{R}%
\right) ,
\end{eqnarray}
{and therefore} Eq. (\ref{algebra1}) can be recast as:
\begin{equation}
\frak{e}_{7(-25)}=\frak{sl}\left( 2,\mathbb{R}\right) \oplus \frak{f}%
_{4}\dotplus (\mathbb{H}_{S}^{\prime }\otimes \frak{J}_{3}^{\prime }(\mathbb{%
O})),  \label{algebra2}
\end{equation}
which implements the maximal \textit{non-symmetric} embedding (whose compact
form is given \textit{e.g.} by Table 15 of \cite{Slansky}; see also \cite
{Patera-McKay}):
\begin{eqnarray}
E_{7(-25)} &\supset &SL(2,\mathbb{R})\times F_{4\left( -52\right) };  \notag
\\
\mathbf{56} &=&\left( \mathbf{4},\mathbf{1}\right) +\left( \mathbf{2},%
\mathbf{26}\right) ;  \notag \\
\mathbf{133} &=&\left( \mathbf{3},\mathbf{1}\right) +\left( \mathbf{1},%
\mathbf{52}\right) +\left( \mathbf{3},\mathbf{26}\right) .
\label{f4embedding}
\end{eqnarray}
We note in passing that, from the branching (\ref{f4embedding}) of $\mathbf{%
56}$, this embedding is relevant for the \textit{maximal} truncation of $%
\mathcal{N}=2$, $D=4$ magical exceptional theory (based on rank-$3$ simple
Jordan algebra $\frak{J}_{3}(\mathbb{O})$) to the smallest cubic $\mathcal{N}%
=2$, $D=4$ model, namely the so-called $T^{3}$ model, the truncation
condition on the vectors (and their field strengths' fluxes, namely electric
and magnetic charges) being given by $\left( \mathbf{2},\mathbf{26}\right)
=0 $.

As the next step, one needs to identify the subalgebra generating the
maximal compact subgroup $K:=E_{6\left( -78\right) }\times U(1)/\mathbb{Z}%
_{3}$ of $E_{7(-25)}$. By considering the manifestly $\frak{f}_{4(-52)}$%
-covariant decomposition of $\frak{e}_{6\left( -78\right) }$ from Tits'
formula (\ref{Tits-formula}):
\begin{equation}
\frak{e}_{6\left( -78\right) }=\mathcal{L}\left( \mathbb{C},\mathbb{O}%
\right) =\text{Der}(\frak{J}_{3}(\mathbb{O}))\dotplus \left( i\otimes \frak{J%
}_{3}^{\prime }(\mathbb{O})\right) ,  \label{e6-78}
\end{equation}
the Lie algebra $\frak{K}$ of $K$ can be identified as the subalgebra of $%
\frak{e}_{6\left( -78\right) }$ defined by picking the only imaginary unit $%
i\in \mathbb{H}_{S}$ which satisfies $i^{2}=-1$ and computing:
\begin{equation}
\frak{K}=ad_{i}\oplus \text{Der}(\frak{J}_{3}(\mathbb{O}))\dotplus \left(
i\otimes \frak{J}_{3}^{\prime }(\mathbb{O})\right) ,  \label{e7-25}
\end{equation}
where $ad_{i}\in \mathbb{H}_{S}$ denotes the \textit{adjoint action} of $i$,
generating the maximal compact subgroup $U(1)$ of $SL(2,\mathbb{R})$. It is
worth remarking that, due to the following property of the Lie product:
\begin{equation}
\lbrack i\otimes j_{1},i\otimes
j_{2}]=-[L_{j_{1}},L_{j_{2}}],~~~\,j_{1},j_{2}\in \frak{J}_{3}^{\prime }(%
\mathbb{O}),
\end{equation}
the multiplication of $\frak{J}_{3}^{\prime }(\mathbb{O})$ by the imaginary
unit $i$ in the last summand of (\ref{e6-78}) and (\ref{e7-25}) is exactly
what is needed to get the compact form of $E_{6(-78)}$ instead of the
(minimally) non-compact real form $E_{6(-26)}$, when exponentiating the
algebra.\newline

As anticipated, by this procedure, inspired by the approach of \cite
{E7magic} and exploiting the methods explained in \cite{IY}, one can
construct the (smallest symplectic) fundamental irrep. $\mathbf{Fund}=\mathbf{56}$ of $E_{7\left( -25\right) }$
reproducing the structure constants of the $\mathbf{Adj}=\mathbf{133}$ irrep. (for whatever basis one chooses for the algebra).

Such an explicit symplectic realization reads as follows:

\begin{equation}
Y_{I}=\left(
\begin{tabular}{c|c|c|c}
$\phi _{I}$ & $\overrightarrow{0}_{27}$ & $0_{27}$ & $\overrightarrow{0}%
_{27} $ \\ \hline
&  &  &  \\[-0.8em]
$\overrightarrow{0}_{27}^{T}$ & $0$ & $\overrightarrow{0}_{27}^{T}$ & $0$ \\
\hline
&  &  &  \\[-0.8em]
$0_{27}$ & $\overrightarrow{0}_{27}$ & $-\phi _{I}^{T}$ & $\overrightarrow{0}%
_{27}$ \\ \hline
&  &  &  \\[-0.8em]
$\overrightarrow{0}_{27}^{T}$ & $0$ & $\overrightarrow{0}_{27}^{T}$ & $0$%
\end{tabular}
\right) ,~I=1,...,78;  \label{Y_I}
\end{equation}
\begin{equation}
Y_{79}=\left(
\begin{tabular}{c|c|c|c}
$\frac{i}{\sqrt{6}}I_{27}$ & $\overrightarrow{0}_{27}$ & $0_{27}$ & $%
\overrightarrow{0}_{27}$ \\ \hline
&  &  &  \\[-0.8em]
$\overrightarrow{0}_{27}^{T}$ & $-i\sqrt{\frac{3}{2}}$ & $\overrightarrow{0}%
_{27}^{T}$ & $0$ \\ \hline
&  &  &  \\[-0.8em]
$0_{27}$ & $\overrightarrow{0}_{27}$ & $-\frac{i}{\sqrt{6}}I_{27}$ & $%
\overrightarrow{0}_{27}$ \\ \hline
&  &  &  \\[-0.8em]
$\overrightarrow{0}_{27}^{T}$ & $0$ & $\overrightarrow{0}_{27}^{T}$ & $i%
\sqrt{\frac{3}{2}}$%
\end{tabular}
\right) ;  \label{Y_79}
\end{equation}
\begin{equation}
Y_{\alpha +79}=\frac{1}{2}\left(
\begin{tabular}{c|c|c|c}
$0_{27}$ & $\overrightarrow{0}_{27}$ & $2iA_{\alpha }$ & $i\sqrt{2}%
\overrightarrow{e}_{\alpha }$ \\ \hline
$\overrightarrow{0}_{27}^{T}$ & $0$ & $i\sqrt{2}\overrightarrow{e}_{\alpha
}^{T}$ & $0$ \\ \hline
&  &  &  \\[-0.8em]
$-2iA_{\alpha }$ & $-i\sqrt{2}\overrightarrow{e}_{\alpha }$ & $0_{27}$ & $%
\overrightarrow{0}_{27}$ \\ \hline
&  &  &  \\[-0.8em]
$-i\sqrt{2}\overrightarrow{e}_{\alpha }^{T}$ & $0$ & $\overrightarrow{0}%
_{27}^{T}$ & $0$%
\end{tabular}
\right) ,~\alpha =1,...,27;  \label{Y_alpha+79}
\end{equation}
\begin{equation}
Y_{\alpha +106}=\frac{1}{2}\left(
\begin{tabular}{c|c|c|c}
$0_{27}$ & $\overrightarrow{0}_{27}$ & $-2A_{\alpha }$ & $\sqrt{2}%
\overrightarrow{e}_{\alpha }$ \\ \hline
&  &  &  \\[-0.8em]
$\overrightarrow{0}_{27}^{T}$ & $0$ & $\sqrt{2}\overrightarrow{e}_{\alpha
}^{T}$ & $0$ \\ \hline
&  &  &  \\[-0.8em]
$-2A_{\alpha }$ & $\sqrt{2}\overrightarrow{e}_{\alpha }$ & $0_{27}$ & $%
\overrightarrow{0}_{27}$ \\ \hline
&  &  &  \\[-0.8em]
$\sqrt{2}\overrightarrow{e}_{\alpha }^{T}$ & $0$ & $\overrightarrow{0}%
_{27}^{T}$ & $0$%
\end{tabular}
\right) ,~\alpha =1,...,27,  \label{Y_alpha+106}
\end{equation}
where $I_{n}$ is the $n\times n$ identity matrix, $0_{27}$ is the $27\times
27$ null matrix, $\overrightarrow{0}_{n}$ is the zero vector in $\mathbb{R}%
^{n}$, and $\overrightarrow{e}_{\alpha }$, $\alpha =1,...,27$, is the
canonical basis of $\mathbb{R}^{27}$ throughout.

The $78$ matrices $\phi _{I}$ realize a subalgebra $\frak{e}_{6\left(
-78\right) }$ in its irreducible representation $\mathbf{Fund}=\mathbf{27}$.
An explicit expression can be found \textit{e.g.} in Sec. 2.1 of \cite
{E7magic}:
\begin{equation}
\phi _{I}=\left\{
\begin{array}{ll}
C_{I} & I=1,\ldots ,52,\cr\tilde{C}_{I-52} & I=53,\ldots ,78,
\end{array}
\right.  \label{phi-split}
\end{equation}
where, in turn, the matrices $C_{I}$ realize a maximal $\frak{f}_{4\left(
-52\right) }$ subalgebra in its $\mathbf{Fund}=\mathbf{26}$ irrep. (see
\textit{e.g.} \cite{F4, E6}).

The 27 matrices $A_\alpha$ are related to the $d$-tensor of $E_6$, as
explained in more detail in the next Subsec.~\ref{sec:d-tensor}.

The first $79$ matrices $Y_{I}$ (\ref{Y_I}) and $Y_{79}$ (\ref{Y_79})
generate the maximal compact subgroup $K$ of $E_{7\left( -25\right) }$ and
are anti-hermitian, whereas the remaining ones $Y_{\alpha +79}$ (\ref
{Y_alpha+79}) and $Y_{\alpha +106}$ (\ref{Y_alpha+106}) generate the
non-compact symmetric coset $E_{7\left( -25\right) }/K$ and they are
hermitian.

By introducing \textit{(cfr.} \cite{E7magic}):
\begin{equation}
\widetilde{I}:=\left(
\begin{tabular}{c|c}
$I_{26}$ & $\overrightarrow{0}_{26}$ \\ \hline
&  \\[-0.8em]
$\overrightarrow{0}_{26}^{T}$ & $-2$%
\end{tabular}
\right) ,
\end{equation}
the two matrices $Y_{106}$ and $Y_{133}$ can be rewritten more explicitly as:
\begin{equation}
Y_{106}=\frac{1}{2}\left(
\begin{tabular}{c|c|c|c}
$0_{27}$ & $\overrightarrow{0}_{27}$ & $-i\sqrt{\frac{2}{3}}\widetilde{I}$ &
$i\sqrt{2}\overrightarrow{e}_{27}$ \\ \hline
&  &  &  \\[-0.8em]
$\overrightarrow{0}_{27}^{T}$ & $0$ & $i\sqrt{2}\overrightarrow{e}_{27}^{T}$
& $0$ \\ \hline
&  &  &  \\[-0.8em]
$i\sqrt{\frac{2}{3}}\widetilde{I}$ & $-i\sqrt{2}\overrightarrow{e}_{27}$ & $%
0_{27}$ & $\overrightarrow{0}_{27}$ \\ \hline
&  &  &  \\[-0.8em]
$-i\sqrt{2}\overrightarrow{e}_{27}^{T}$ & $0$ & $\overrightarrow{0}_{27}^{T}$
& $0$%
\end{tabular}
\right) ,  \label{Y_106}
\end{equation}
\begin{equation}
Y_{133}=\frac{1}{2}\left(
\begin{tabular}{c|c|c|c}
$0_{27}$ & $\overrightarrow{0}_{27}$ & $\sqrt{\frac{2}{3}}\widetilde{I}$ & $%
\sqrt{2}\overrightarrow{e}_{27}$ \\ \hline
&  &  &  \\[-0.8em]
$\overrightarrow{0}_{27}^{T}$ & $0$ & $\sqrt{2}\overrightarrow{e}_{27}^{T}$
& $0$ \\ \hline
&  &  &  \\[-0.8em]
$\sqrt{\frac{2}{3}}\widetilde{I}$ & $\sqrt{2}\overrightarrow{e}_{27}$ & $%
0_{27}$ & $\overrightarrow{0}_{27}$ \\ \hline
&  &  &  \\[-0.8em]
$\sqrt{2}\overrightarrow{e}_{27}^{T}$ & $0$ & $\overrightarrow{0}_{27}^{T}$
& $0$%
\end{tabular}
\right) .  \label{Y_133}
\end{equation}
Together with $Y_{79}$ ($U(1)$ generator), $Y_{106}$ and $Y_{133}$ generate
an $SL\left( 2,\mathbb{R}\right) $ subgroup, corresponding to the one
appearing in Eqs. (\ref{algebra2}) and (\ref{f4embedding}).

\subsection{\label{sec:d-tensor}The matrices $A_{\protect\alpha}$ and the $d$-tensor of the $\mathbf{27}$ of $E_{6\left( -78\right)}$}

By choosing a basis $\{j_{a}\}_{a=1,...,26}$ of $\frak{J}_{3}^{\prime }(%
\mathbb{O})$ normalized as $\langle j_{a},j_{b}\rangle =2\delta _{ab}$, a
completion to a basis for $\frak{J}_{3}(\mathbb{O})$ can be obtained by
adding $j_{27}=\sqrt{\frac{2}{3}}I_{3}$. The $A_{\alpha }$'s are $27\times
27 $ symmetric matrices representing, by means of the linear isomorphism $%
\frak{J}_{3}(\mathbb{O})\simeq \mathbb{R}^{27},\ \ j_{\alpha }\mapsto \vec{e}%
_{\alpha }$, the action $\rhd $ of $\frak{J}_{3}(\mathbb{O})$ on $\frak{J}%
_{3}(\mathbb{O})$ itself. The components of $A_{\alpha }$, explicitly
computed in \cite{E7magic}, satisfy the following relation \cite{freudenthal}%
:
\begin{equation}
(A_{\alpha })_{\ \gamma }^{\beta }=\frac{1}{2}\text{Tr}\left( (j_{\alpha
}\rhd j_{\gamma })\circ j_{\beta }\right) =\frac{3}{2}\,Det(j_{\alpha
},j_{\gamma },j_{\beta })=:\frac{1}{\sqrt{2}}\,d_{\alpha \gamma \beta },
\label{d-tensor}
\end{equation}
where $d_{\alpha \gamma \beta }=d_{\left( \alpha \gamma \beta \right) }$ is
the totally symmetric rank-$3$ invariant $d$-tensor of the $\mathbf{27}$ of
of $E_{6\left( -78\right) }$, with a normalization suitable to match $%
Det(j_{\alpha },j_{\gamma },j_{\beta })$ given by (\ref{determinant}) (see
below). We point out that the result (\ref{d-tensor}) does not depend on the
choice of the basis $\{j_{\alpha }\}$. Thus, the expressions of $Y_{\alpha
+79}$ (\ref{Y_alpha+79}) and of $Y_{\alpha +106}$ (\ref{Y_alpha+106})
exhibit the maximal manifest compact $\left[ (E_{6\left( -78\right) }\times U(1))/\mathbb{Z}_3\right] $-covariance.
However, whenever the choice of the basis $\{j_{\alpha }\}$ is exploited in order to distinguish the identity matrix
from the traceless ones, the $d_{\alpha \beta \gamma }$ of $E_{6}$ has a
maximal manifestly $F_{4\left( -52\right) }$-invariance only. This also
holds for the expressions of the $Y_{I}$ (\ref{Y_I}), which are manifestly $%
F_{4\left( -52\right) }$-covariant only, due to the splitting (\ref
{phi-split}). Notice that the full $\left[ (E_{6\left( -78\right) }\times U(1))/\mathbb{Z}_3 \right] $-covariance
can be recovered simply by picking a generic basis for
the Jordan algebra.

A manifestly $\left[ SU\left( 6\right) \times SU\left( 2\right) \right] $%
-invariant expression of the $d$-tensor of the $\mathbf{27}$ of $E_{6\left(
-78\right) }$ has been constructed in \cite{BMP-1}. On the other hand, $d$%
-tensors for the non-compact real forms of $E_{6}$ have been more
extensively considered in the literature, e.g. due to their appearance in
the general form of the holomorphic prepotential $F$ of cubic special
K\"{a}hler geometry (see \textit{e.g.} \cite{dWVVP}). For
instance, in \cite{fgk} the $d$-tensors of $E_{6\left( 6\right) }$ (split)
and $E_{6\left( -26\right) }$ (minimally non-compact) real forms have been
explicitly constructed, with $USp\left( 8\right) $ and $USp\left( 6,2\right)
$ maximal manifest invariance, respectively. By denoting with $G_{6}$ the $U$%
-duality\footnote{%
Here $U$-duality is referred to as the ``continuous'' symmetries of \cite
{CJ-1}. Their discrete versions are the $U$-duality non-perturbative string
theory symmetries introduced by Hull and Townsend \cite{HT-1}.} group of
chiral supergravity theories with $8$ supersymmetries in $D=6$ space-time
dimensions, and considering all $U$-duality groups $G_{5}$ of $\mathcal{N}=2$%
, $D=5$ supergravity theories with symmetric (vector multiplets') scalar
manifold, manifestly $\left[ G_{6}\times SO\left( 1,1\right) \right] $%
-invariant expressions of the $G_{5}$-invariant $d$-tensor have been derived
\textit{e.g.} in \cite
{dWVVP,dWVP-cubic,AFMT-1,Ferrara-Maldacena,FG-D=5,ICL-1}.

A necessary remark on the consistence of normalizations is in order. As a
consequence of the choice (\ref{Ynorm}) for the normalization of the
matrices $Y_{A}$ discussed in the next Subsec.~\ref{Properties-of-Y}, the
components $(A_{\alpha })_{\ \gamma }^{\beta }:=A_{\alpha \beta \gamma }$
are normalized as:
\begin{equation}
A_{\alpha \beta \gamma }A^{\eta \beta \gamma }=5\delta _{\alpha }^{\eta }.
\label{AA-norm}
\end{equation}
This is consistent with the normalization of the $d$-tensor (of $E_{6\left(
-26\right) }$) given by the following expression of the K\"{a}hler-invariant
($\left( X^{0}\right) ^{2}$-rescaled) holomorphic prepotential function
characterizing special K\"{a}hler geometry (see \textit{e.g.} \cite
{Strominger-SKG,dWVVP,N=2-Big}):
\begin{equation}
f\left( z\right) :=\frac{1}{3!}d_{\alpha \beta \gamma }z^{\alpha }z^{\beta
}z^{\gamma },  \label{f}
\end{equation}
adopted \textit{e.g.} in \cite{adfl}; in general, $\alpha =1,...,n_{V}$,
where $n_{V}$ denotes the number of Abelian vector multiplets coupled to the
supergravity multiplet. Indeed, within the notation conventions adopted in
\cite{CFM-1}, one can compute that (see also \cite{CFM-2} and \cite{Raju-1}):
\begin{equation}
d_{\alpha \beta \gamma }d^{\eta \beta \gamma }=\left( q+2\right) \delta
_{\alpha }^{\eta }.  \label{dd-norm}
\end{equation}
For all the models reported in Table \ref{scalarmnf} below but the $T^{3}$ model, $q$ can be
defined as:
\begin{equation}
q\equiv \text{dim}_{\mathbb{R}}\mathbb{A},  \label{def-q}
\end{equation}
where $\mathbb{A}$ denotes the division algebra on which the corresponding
rank-$3$ simple Jordan algebra $\frak{J}_{3}(\mathbb{A})$ is constructed ($%
q=8$, $4$, $2$, $1$ for $\mathbb{A}=\mathbb{O}$, $\mathbb{H}$, $\mathbb{C}$,
$\mathbb{R}$, respectively). Furthermore, as observed in \cite{LM-1}, in
general $q$ is related to the \textit{inverse Coxeter number} $\mathbf{%
\lambda }$ by the relation:
\begin{eqnarray}
\mathbf{\lambda } &=&-\frac{2}{q+2},~q=0,1,2,4,8;  \label{jjj} \\
\mathbf{\lambda } &=&-\frac{1}{q+1},~q=-2/3~\text{(}T^{3}~\text{model)}.
\end{eqnarray}
The case $q=0$ in (\ref{jjj}) corresponds to the \textit{triality symmetric}
so-called $\mathcal{N}=2$ $STU$ model \cite{stu}, based on $\frak{J}_{3}=%
\mathbb{R}\oplus \mathbf{\Gamma }_{1,1}\sim \mathbb{R}\oplus \mathbb{R}%
\oplus \mathbb{R}$; however, since the corresponding $U$-duality group $%
G_{4} $ is \textit{semi-simple}, it will not be considered in the present
investigation.

Coming back to the previous reasoning, by plugging $q=8$ (corresponding to
the octonionic theory considered above) into (\ref{dd-norm}), one achieves
the following result:
\begin{equation}
q=8:d_{\alpha \beta \gamma }d^{\eta \beta \gamma }=10\delta _{\alpha }^{\eta
},
\end{equation}
which matches (\ref{AA-norm}) when taking (\ref{d-tensor}) into account, and
assuming for the $d$-tensor of $E_{6\left( -78\right) }$ the same
normalization of the $d$-tensor of $E_{6\left( -26\right) }$.

\subsection{\label{Properties-of-Y}Properties of the Matrices $Y_A$}

The $Y_{A}$'s are orthonormalized (with signature $(-^{79},+^{54})$) with
respect to the product:
\begin{equation}
\langle Y,Y^{\prime }\rangle _{\mathbf{56}}:=\frac{1}{12}\text{Tr}%
(YY^{\prime }).  \label{Ynorm}
\end{equation}
This normalization guarantees that the period of the maximal torus in the $%
E_6$ subgroup equals $4 \pi$, which is the standard choice for the period of
the spin representations of the orthogonal subgroups \cite{F4, E6}.

Furthermore, the complete symmetry of the $d$-tensor implies the matrices $%
Y_{A}$ ($A=1,...,133$) given by the expressions (\ref{Y_I})-(\ref
{Y_alpha+106}) to be \textit{symplectic} with respect to the canonical
symplectic form:
\begin{equation}
\Omega :=\left(
\begin{array}{cc}
0_{28} & -I_{28} \\
I_{28} & 0_{28}
\end{array}
\right) ,  \label{Omega}
\end{equation}
namely (in a block-wise notation, and suppressing the index $A$):
\begin{equation}
Y:=\left(
\begin{array}{cc}
A & B \\
C & D
\end{array}
\right) \in \frak{sp}\left( 56,\mathbb{C}\right) \Leftrightarrow \Omega
Y+Y^{T}\Omega =0\Leftrightarrow \left\{
\begin{array}{rl}
A^{T} & =-D \\
B^{T} & =B \\
\quad C^{T} & =C
\end{array}
\right.
\end{equation}

Actually, it holds that:
\begin{equation}
Y\in \frak{usp}\left( 28,28\right) .  \label{usp(28,28)}
\end{equation}
In order to show this, let us introduce:
\begin{equation}
\mathcal{H}:=\left(
\begin{array}{cc}
I_{28} & 0_{28} \\
0_{28} & -I_{28}
\end{array}
\right) ,
\end{equation}
and recall the infinitesimal condition:
\begin{equation}
Y\in \frak{u}(28,28)\Leftrightarrow \mathcal{H}Y+Y^{\dagger }\mathcal{H}%
=0\Leftrightarrow \left\{
\begin{array}{rl}
A^{\dagger } & =-A \\
B^{\dagger } & =C \\
D^{\dagger } & =-D
\end{array}
\right. .
\end{equation}
Thus, by means of the isomorphism:
\begin{equation}
\frak{sp}\left( 2n,\mathbb{R}\right) \sim \frak{usp}\left( n,n\right) \equiv
\frak{sp}\left( 2n,\mathbb{C}\right) \cap \frak{u}\left( n,n\right) ,
\label{iso-1}
\end{equation}
it follows that:
\begin{equation}
Y\in \frak{usp}\left( 28,28\right) \Leftrightarrow \left\{
\begin{array}{rcl}
A & =-A^{\dagger } & =\overline{D}; \\
C & =B^{\dagger } & =\overline{B}.
\end{array}
\right.
\end{equation}
It should be noted that, when considering $n$ vector fields in presence of
scalar fields, the isomorphism (\ref{iso-1}) has been exploited by Gaillard
and Zumino in \cite{GZ} for the study of the generalization and non-compact
nature of the electric-magnetic symmetry, naturally yielding in $D=4$ a
manifestly $USp\left( n,n\right) $-covariant basis of
self-dual/anti-self-dual vector 2-form field strengths, rather than an $%
Sp\left( 2n,\mathbb{R}\right) $-covariant one; see also \textit{e.g.} the
re-elaboration of such a treatment presented in \cite{N=2-Big}.

Let us analyze the properties of the matrices $Y_{A}$ (\ref{Y_I})-(\ref
{Y_alpha+106}) (following the notation of \cite{E7magic}):

\begin{enumerate}
\item  $Y_{I}$ (\ref{Y_I}) with $I=1,...,52$. According to (\ref{phi-split}%
), $\phi _{I}=C_{I}$. Up to a change of basis of the Jordan algebra, the
matrices $C_{I}$ are given in \cite{F4} (including the \textit{Mathematica}
routine used for their computation). As mentioned before, the $C_{I}$'s
realize a maximal $\frak{f}_{4\left( -52\right) }$ subalgebra in its
irreducible representation $\mathbf{Fund}=\mathbf{26}$. In turn, this is
embedded into the algebra $\frak{e}_{6\left( -78\right) }$ (maximal compact
subalgebra of $\frak{e}_{7\left( -25\right) }$) in its $\mathbf{Fund}=%
\mathbf{27}$ irrep., through the addition of an extra 27th row and column of
$0$'s, according to the maximal and symmetric embedding : $E_{6}\supset
F_{4} $, $\mathbf{27}=\mathbf{26}+\mathbf{1}$. The symmetry properties are:
\begin{equation}
C_{I}=-C_{I}^{T},\quad \overline{C_{I}}=C_{I}\Longrightarrow
Y_{I}=-Y_{I}^{\dagger }\in \frak{usp}\left( 28,28\right) .  \label{1}
\end{equation}

\item  $Y_{I}$ (\ref{Y_I}) with $I=53,...,78$. According to (\ref{phi-split}%
), $\Phi _{I}=\tilde{C}_{I-52}$, as computed in \cite{E6}, where the \textit{%
Mathematica} routine to generate them is given, as well. The fact that the $%
\tilde{C}$'s are purely imaginary is a consequence of the presence of the
factor $i$ in the last summand of Eq. (\ref{e6-78}); they are defined in
terms of the action (\ref{jaction}) applied to the traceless part $\frak{J}%
_{3}^{\prime }\left( \mathbb{O}\right) $ of the Jordan algebra. In turn,
such an action of the Jordan algebra on itself is the one entering the cubic
form and hence in the definition (\ref{d-tensor}) of the matrices $A_{\alpha
}$'s, implying that the $\tilde{C}_{I-52}$ coincide with the first $26$
components of $A_{\alpha }$, apart from an overall $i$. The symmetry
properties are:
\begin{equation}
\tilde{C}_{I-52}={}\tilde{C}_{I-52}^{T},\quad \tilde{C}_{I-52}^{\dagger }=-%
\tilde{C}_{I-52}\Longrightarrow Y_{I}=-Y_{I}^{\dagger }\in \frak{usp}\left(
28,28\right) .  \label{2}
\end{equation}

\item  $Y_{79}$ (\ref{Y_79}). It generates a $U\left( 1\right) $ subgroup,
corresponding to the compact Cartan of the $SL\left( 2,\mathbb{R}\right) $
factor group, appearing in Eqs. (\ref{algebra2}) and (\ref{f4embedding}).
The symmetry properties are:
\begin{equation}
Y_{79}=Y_{79}^{T},\quad Y_{79}^{\dagger }=-Y_{79}\Longrightarrow Y_{79}\in
\frak{usp}\left( 28,28\right) .  \label{3}
\end{equation}

\item  $Y_{I}$ with $I=80,...,106$, \textit{i.e.} $Y_{\alpha +79}$ (\ref
{Y_alpha+79}). The symmetry properties read as follows:
\begin{equation}
A_{\alpha }={}A_{\alpha }^{T},~A_{\alpha }^{\dagger }=A_{\alpha
}\Longrightarrow Y_{\alpha +79}^{\dagger }=Y_{\alpha +79},~Y_{\alpha
+79}=-{}Y_{\alpha +79}^{T},~~Y_{\alpha +79}\in \frak{usp}(28,28).  \label{4}
\end{equation}

\item  $Y_{I}$ with $I=107,...,133$, \textit{i.e.} $Y_{\alpha +106}$ (\ref
{Y_alpha+106}). The symmetry properties read as follows:
\begin{equation}
Y_{\alpha +106}^{\dagger }=Y_{\alpha +106},~Y_{\alpha +106}=-Y_{\alpha
+106}^{T},~~Y_{\alpha +106}\in \frak{usp}(28,28).  \label{5}
\end{equation}
\end{enumerate}

Thus, (\ref{usp(28,28)}) results from (\ref{1})-(\ref{5}).

As elucidated in the next section, the matrices $Y_{I}$, $I=80,\ldots ,133$
given by (\ref{Y_alpha+79}) and (\ref{Y_alpha+106}) are the Hermitian
generators of the symmetric maximal non-compact (special K\"{a}hler)
Riemannian coset (\ref{M-call}), which is the classical vector multiplets'
scalar manifold of the magical $\mathcal{N}=2$, $D=4$ Maxwell-Einstein
supergravity theory based on $\frak{J}_{3}\left( \mathbb{O}\right) $ \cite
{GST}. As given by Eq. (\ref{d-tensor}), the $27$ matrices $A_{\alpha }$ are
directly related to the invariant $d$-tensor of the $\mathbf{27}$ irrep. of $%
E_{6\left( -78\right) }$; they have been explicitly constructed in \cite
{E7magic}, to which the reader is addressed for further detail.

\section{\label{manifest}Manifestly $\mathbf{\left[ (E_{6\left( -78\right) }\times U(1))/\mathbb{Z}_3 \right]}$-covariant Coset Construction}

The quotient manifold $\mathcal{M}$ (\ref{M-call}) has rank $3$; this means
that the maximal dimension of the intersection between a Cartan subalgebra
of $E_{7\left( -25\right) }$ and the generators of $\mathcal{M}$ itself is $%
3 $. From the results reported above, the $3$ generators of a Cartan
subalgebra of $\mathcal{M}$ can be chosen to be the diagonal generators of
the Jordan algebra $\frak{J}_{3}\left( \mathbb{O}\right) $ itself, namely $%
Y_{123}$, $Y_{132}$ and $Y_{133}$.

The coset $\mathcal{M}$ (\ref{M-call}) is generated by the matrices $%
Y_{79+I} $, (\ref{Y_alpha+79}) and (\ref{Y_alpha+106}) with $I=1,\ldots ,54$%
. Through the exponential mapping, it can be defined as follows:
\begin{equation}
\mathcal{M}:=\exp \left( \sum_{\alpha =1}^{27}x_{\alpha }Y_{106+\alpha
}+y_{\alpha }Y_{79+\alpha }\right) ,  \label{coset1}
\end{equation}
with $x_{\alpha }$, $y_{\alpha }$, $\alpha =1,\ldots 27$, real parameters.

From the commutation relations of the matrices $Y$'s, which can be easily
computed by means of the Mathematica program provided in \cite{E7magic}, it
holds that:
\begin{equation}
\lbrack Y_{51+\alpha },Y_{106}]=-\sqrt{\frac{2}{3}}Y_{106+\alpha },\quad
\lbrack Y_{51+\alpha },Y_{133}]=\sqrt{\frac{2}{3}}Y_{79+\alpha }.
\end{equation}
The generators of $\frak{e}_{6\left( -78\right) }$ which are not in $\frak{f}%
_{4\left( -52\right) }$ mix the matrices $Y_{79+\alpha }$ with the $%
Y_{106+\alpha }$. Therefore, in order to make the complex structure of $%
\mathcal{M}$ manifest, it is necessary to introduce the following complex
linear combinations of the matrices:
\begin{eqnarray}
\zeta _{\alpha }:= &&\frac{1}{\sqrt{2}}\left( Y_{79+\alpha
}+i\;Y_{106+\alpha }\right) , \\
\bar{\zeta}_{\alpha }:= &&\frac{1}{\sqrt{2}}\left( Y_{79+\alpha
}-i\;Y_{106+\alpha }\right) .  \notag
\end{eqnarray}
This hints for the complex linear combinations of the parameters:
\begin{eqnarray}
z_{\alpha }:= &&\frac{1}{\sqrt{2}}(y_{\alpha }+i\,x_{\alpha }),
\label{z-alpha} \\
\bar{z}_{\alpha }:= &&\frac{1}{\sqrt{2}}(y_{\alpha }-i\,x_{\alpha }),  \notag
\label{z-alpha-bar}
\end{eqnarray}
which allows one to rewrite (\ref{coset1}) as:
\begin{equation}
\mathcal{M}:=\exp \left( \sum_{\alpha =1}^{27}\bar{z}_{\alpha }\zeta
_{\alpha }+z_{\alpha }\bar{\zeta}_{\alpha }\right) .  \label{coset2}
\end{equation}
By introducing the $27$ dimensional complex vector:
\begin{equation}
z:={\sum_{\alpha =1}^{27}}z_{\alpha }\vec{e}_{\alpha },  \label{z-vector}
\end{equation}
and the $28\times 28$ matrix:
\begin{equation}
\mathcal{A}:=\left(
\begin{array}{c|c}
\displaystyle{-\sqrt{2}{\sum_{\alpha =1}^{27}}\bar{z}_{\alpha }A_{\alpha }}
& z \\[3ex] \hline
&  \\[-1.6ex]
\;z^{T} & 0
\end{array}
\right) ,  \label{dcoset}
\end{equation}
Eq. (\ref{coset2}) enjoys the simple form:
\begin{equation}
\mathcal{M}:=\exp \left(
\begin{array}{c|c}
0 & \mathcal{A} \\ \hline
&  \\[-1.6ex]
\mathcal{A}^{\dagger } & 0
\end{array}
\right) =\left(
\begin{array}{c|c}
\text{Ch}(\sqrt{\mathcal{AA}^{\dagger }}) & \displaystyle{\mathcal{A} \frac{\text{Sh}(\sqrt{\mathcal{A}^{\dagger }\mathcal{A}})}{\sqrt{\mathcal{A}%
^{\dagger }\mathcal{A}}}} \\[2.5ex] \hline
&  \\[-1.2ex]
\displaystyle{\frac{\text{Sh}(\sqrt{\mathcal{AA}^{\dagger }})}{\sqrt{\mathcal{AA}^{\dagger }}}\mathcal{A}^{\dagger }} & \text{Ch}(\sqrt{\mathcal{A}%
^{\dagger }\mathcal{A}})
\end{array}
\right) .  \label{dcoset2}
\end{equation}
This is a Hermitian matrix, of the same form as the finite coset
representative worked out \cite{DeWitNicolai} for the \textit{split} (%
\textit{i.e.} maximally non-compact) counterpart
\begin{equation}
\mathcal{M}_{\mathcal{N}=8}=\frac{E_{7\left( 7\right) }}{SU\left( 8\right) /%
\mathbb{Z}_{2}},  \label{M-call-N=8}
\end{equation}
which is the scalar manifold of \textit{maximal} $\mathcal{N}=8$, $D=4$
supergravity, associated to $\frak{J}_{3}\left( \mathbb{O}_{S}\right) $. On
the other hand, as a consequence of (\ref{usp(28,28)}), $\mathcal{M}$ also
is an element of $USp\left( 28,28\right) $, whereas $\mathcal{M}_{\mathcal{N}%
=8}$ is real.

By using the machinery of \textit{special K\"{a}hler geometry} (see \textit{%
e.g.} \cite{Strominger-SKG,dWVVP,N=2-Big}), the
symplectic sections defining the \textit{symplectic frame} associated to the
coset parametrization introduced above can be directly read from (\ref
{dcoset})-(\ref{dcoset2}):
\begin{equation}
\mathcal{M}=:\left(
\begin{array}{c|c}
u_{i}^{\Lambda }\left( z,\overline{z}\right) & v_{i\Lambda }\left( z,%
\overline{z}\right) \\[0.8ex] \hline
&  \\[-1ex]
v^{i\Lambda }\left( z,\overline{z}\right) & u_{\Lambda }^{i}\left( z,%
\overline{z}\right)
\end{array}
\right) ,  \label{M-call-par}
\end{equation}
where the symplectic index $\Lambda =0,1,...27$ (with $0$ pertaining to the $%
\mathcal{N}=2$, $D=4$ graviphoton), and $i=\overline{\alpha },28$. Thus, the
symplectic sections read (see \textit{e.g.} \cite{CDF-rev,N=2-Big}; subscript ``$28$'' omitted):
\begin{eqnarray}
f_{i}^{\Lambda } &:&=\frac{1}{\sqrt{2}}\left( u+v\right) _{i}^{\Lambda
}=\left( \overline{f}_{\overline{\alpha }}^{\Lambda },f^{\Lambda }\right)
:=\left( \overline{\mathcal{D}}_{\overline{\alpha }}\overline{L}^{\Lambda
},L^{\Lambda }\right) =\exp \left( \frac{1}{2}K\right) \left( \overline{%
\mathcal{D}}_{\overline{\alpha }}\overline{X}^{\Lambda },X^{\Lambda }\right)
;  \label{f-sect} \\
h_{i\Lambda } &:&=-\frac{i}{\sqrt{2}}\left( u-v\right) _{i\Lambda }=\left(
\overline{h}_{\overline{\alpha }\mid \Lambda },h_{\Lambda }\right) :=\left(
\overline{\mathcal{D}}_{\overline{\alpha }}\overline{M}_{\Lambda
},M_{\Lambda }\right) =\exp \left( \frac{1}{2}K\right) \left( \overline{%
\mathcal{D}}_{\overline{\alpha }}\overline{F}_{\Lambda },F_{\Lambda }\right)
,  \label{h-sect}
\end{eqnarray}
where $\mathcal{D}$ is the K\"{a}hler-covariant differential operator,
\begin{equation}
\mathcal{V}:=\left( L^{\Lambda },M_{\Lambda }\right) ^{T}=\exp \left( \frac{1%
}{2}K\right) \left( X^{\Lambda },F_{\Lambda }\right) ^{T}  \label{V-call}
\end{equation}
is the symplectic vector of K\"{a}hler-covariantly holomorphic sections, and
\begin{equation}
K:=-\ln \left[ i\left( \overline{X}^{\Lambda }F_{\Lambda }-X^{\Lambda }%
\overline{F}_{\Lambda }\right) \right]  \label{K}
\end{equation}
is the K\"{a}hler potential determining the corresponding geometry.

As announced, a key feature of (\ref{dcoset}) is that the matrix $\mathcal{A}
$, generating the coset $\mathcal{M}$ (\ref{M-call}) through (\ref{dcoset2}%
), is written in terms of the invariant rank-$3$ $d$-tensor of the $\mathbf{%
27}$ fundamental irrep. of $E_{6(-78)}$, thus yielding a formalism with
manifest $\left[ (E_{6\left( -78\right) }\times U(1))/\mathbb{Z}_3\right]$-covariance,
which is the maximal compact possible symmetry of the framework under
consideration. Within such a parametrization, the complex scalar fields of
the corresponding $\mathcal{N}=2$ magic theory, coordinatizing $\mathcal{M}
$ (\ref{M-call}), are defined by (\ref{z-alpha}), and summarized in vector
notation by (\ref{z-vector}).\smallskip

Furthermore, attention should be paid not to confuse this symplectic frame
with the so-called \textit{``}$4D/5D$\textit{\ special coordinates'' }
symplectic frame (see \textit{e.g.} \cite{CFM-1}), in
which the holomorphic prepotential function $F$ exists and it is given by ($%
\left( X^{0}\right) ^{2}$ times) Eq. (\ref{f}). Indeed, as commented below, $%
F$ simply does not exist in the symplectic frame under consideration
(namely, $2F=X^{\Lambda }F_{\Lambda }=0$ \cite{CDFVP}), and the $d$-tensor
of the $\mathbf{27}$ of $E_{6\left( -26\right) }$ (appearing in (\ref{f}))
is different from the $d$-tensor of $\mathbf{27}$ of $E_{6\left( -78\right)
} $, appearing in the treatment of Sec. \ref{sec:56} and of the present
section; such a difference is evident \textit{e.g.} when considering a
manifestly $\left[ G_{6}\times SO\left( 1,1\right) \right] $-invariant
formalism, as done \textit{e.g.} in \cite{FG-D=5} and in \cite{fgk}.

As mentioned in Sec. \ref{sec:56}, by exploiting the expressions (\ref{Y_106}%
) and (\ref{Y_133}) of the non compact generators of the relevant $\frak{sl}%
\left( 2,\mathbb{R}\right) $ subalgebra, the maximal manifest $\left[
(E_{6\left( -78\right) }\times U(1))/\mathbb{Z}_3\right] $-covariance can be broken down to
a manifest $\left[ F_{4\left( -52\right) }\times U(1)\right] $-covariance
(recall the maximal symmetric embedding (\ref{algebra2})-(\ref{f4embedding}%
)), in which (\ref{dcoset}) becomes:
\begin{equation}
\mathcal{A}=\left(
\begin{array}{c|c}
\frac{1}{\sqrt{3}}\bar{z}_{27}\tilde{I}-\sqrt{2}\displaystyle{\sum_{a=1}^{26}%
}\bar{z}_{a}A_{a} & z \\[3ex] \hline
&  \\[-1.6ex]
\;z^{T} & 0
\end{array}
\right) .  \label{df4}
\end{equation}
We note in passing that $F_{4\left( -52\right) }$ is particularly relevant,
because it contains all the compact generators of $USp\left( 6,2\right) $, which
is the maximal (non-compact) manifest covariance exhibited by the $d$-tensor
of the $\mathbf{27}$ irrep. of $E_{6\left( -26\right) }$ constructed in \cite
{fgk}:
\begin{equation}
USp\left( 6,2\right) \cap USp\left( 8\right) \cap F_{4\left( -52\right)
}=USp\left( 6\right) \times USp\left( 2\right) \sim USp\left( 6\right)
\times SU\left( 2\right) =\text{mcs}\left[ USp\left( 6,2\right) \right] .
\end{equation}

\subsection{\label{Remarks}Remarks}

In order to gain more insight on the parametrization under consideration, it
is useful to compare the infinitesimal element of the coset $\mathcal{M}$ (%
\ref{M-call}), given by the $28\times 28$ matrix (recall (\ref{dcoset2}) and
(\ref{dcoset})):
\begin{equation}
\ln \mathcal{M}:=\left(
\begin{array}{c|c}
0 & \mathcal{A} \\ \hline
&  \\[-1.6ex]
\mathcal{A}^{\dagger } & 0
\end{array}
\right) ,  \label{ln-M-call}
\end{equation}
with an analogue expression, given by Eq. (6) of \cite{adfl}, which we
recall here for ease of comparison:
\begin{equation}
\mathcal{B}:=\left(
\begin{tabular}{c|c|c|c}
$0_{\beta }^{\alpha }$ & $-t^{\prime \alpha }$ & $d^{\alpha \beta \gamma
}t_{\gamma }$ & $0^{\alpha }$ \\ \hline
$-t_{\beta }$ & $0$ & $0_{\beta }$ & $0$ \\ \hline
$d_{\alpha \beta \gamma }t^{\prime \gamma }$ & $0_{\alpha }$ & $0_{\alpha
}^{\beta }$ & $t_{\alpha }$ \\ \hline
$0_{\beta }$ & $0$ & $t^{\prime \beta }$ & $0$%
\end{tabular}
\right) .  \label{dcoset-B}
\end{equation}
Following the treatment of \cite{adfl}, $\mathcal{B}$ is a real $28\times 28$
matrix depending on $27+27=54$ parameters, parametrizing the generators of
the maximal symmetric non-compact pseudo-Riemannian rank-$3$ coset
\begin{equation}
\widehat{\mathcal{M}}:=\frac{E_{7\left( 7\right) }}{E_{6\left( 6\right)
}\times SO\left( 1,1\right) },  \label{M-call-hat}
\end{equation}
with signature $(-^{27},+^{27})$; in this case, the $d$-tensor appearing in (%
\ref{dcoset-B}) is the one pertaining to the $\mathbf{27}$ (or $\mathbf{27}%
^{\prime }$) irrep. of the split non-compact real form $E_{6\left( 6\right)
} $. On the other hand, by suitably replacing this latter by the $d$-tensor
pertaining to the $\mathbf{27}$ (or $\mathbf{27}^{\prime }$) irrep. of the
minimally non-compact real form $E_{6\left( -26\right) }$, the matrix $%
\mathcal{B}$ (\ref{dcoset-B}) can be regarded as parametrizing the
generators of the maximal symmetric non-compact pseudo-Riemannian rank-$3 $
coset
\begin{equation}
\widetilde{\mathcal{M}}:=\frac{E_{7\left( -25\right) }}{E_{6\left(
-26\right) }\times SO\left( 1,1\right)},  \label{M-call-tilde}
\end{equation}
with signature $(-^{43},+^{11})$; this pseudo-Riemannian counterpart of (\ref
{M-call}) can also be regarded as the classical vector multiplets' scalar
manifold of the magical octonionic $\mathcal{N}=2$ Maxwell-Einstein
supergravity theory in $D=\left( 4,0\right) $ dimensions, obtained from its $%
D=\left( 4,1\right) $ uplift by timelike Kaluza-Klein reduction (see \textit{%
e.g.} \cite{Trig-Berg}). Clearly, also other
interpretations of $\mathcal{B}$ (\ref{dcoset-B}) are possible, within the
maximal (symmetric) embeddings of (non-compact, real forms of) $E_{6}\times
U(1)$ into (non-compact, real forms of) $E_{7}$ (see \textit{e.g.} \cite
{Gilmore}), but they are not relevant for the present
investigation. Notice that in the above expressions (\ref
{M-call-hat}) for $\widehat{\mathcal{M}}$ and (\ref{M-call-tilde}) for $%
\widetilde{\mathcal{M}}$ the issue of the presence of finite or discrete
factors is not taken into account.

We will now relate the matrix $\mathcal{B}$ (\ref{dcoset-B}) (which, within
the interpretation (\ref{M-call-tilde}), provides a manifestly $\left[
E_{6\left( -26\right) }\times SO(1,1)\right] $-covariant parametrization of
the generators of the coset $\widetilde{\mathcal{M}}$) to the matrix $\ln
\mathcal{M}$ (\ref{ln-M-call}) (which provides a manifestly $\left[
(E_{6\left( -78\right) }\times U(1))/\mathbb{Z}_3\right] $-covariant parametrization of the
generators of the coset $\mathcal{M}$ (\ref{M-call})).

\begin{enumerate}
\item  We start and move the vectors $t^{\prime }$ and $t$ from the diagonal
blocks of $\mathcal{B}$ to the off-diagonal ones. In order to achieve this,
a symplectic automorphism generated by the following matrix has to be
performed:
\begin{equation}
\mathcal{S}:=\left(
\begin{tabular}{c|c|c|c}
$I_{27}$ & $0$ & $0$ & $0$ \\ \hline
$0$ & $0$ & $0$ & $-1$ \\ \hline
$0$ & $0$ & $I_{27}$ & $0$ \\ \hline
$0$ & $1$ & $0$ & $0$%
\end{tabular}
\right) .  \label{S-call}
\end{equation}
Thus, it follows that ($\mathcal{S}^{T}\Omega \mathcal{S}=\Omega $)
\begin{equation}
\mathcal{SBS}^{-1}=\left(
\begin{tabular}{c|c|c|c}
$0_{\beta }^{\alpha }$ & $0^{\alpha }$ & $d^{\alpha \beta \gamma }t_{\gamma
} $ & $-t^{\prime \alpha }$ \\ \hline
$0_{\beta }$ & $0$ & $-t_{\beta }^{\prime }$ & $0$ \\ \hline
$d_{\alpha \beta \gamma }t^{\prime \gamma }$ & $-t_{\alpha }$ & $0_{\alpha
}^{\beta }$ & $0_{\alpha }$ \\ \hline
$-t_{\beta }$ & $0$ & $0_{\beta }$ & $0$%
\end{tabular}
\right)
\end{equation}

\item  Then, it is necessary to make the following identification between
the $27$-dimensional vectors $t$, $t^{\prime }$ of $\mathcal{B}$ (\ref
{dcoset-B}) and $z$, $\overline{z}$ defined by (\ref{z-alpha}) and (\ref
{z-vector}):
\begin{equation}
\left\{
\begin{array}{l}
t=-\overline{z}; \\
t^{\prime }=-z.
\end{array}
\right.
\end{equation}

\item  By recalling the normalization of the $d$-tensor given by (\ref
{d-tensor}), it thus follows that $\mathcal{B}$ can be transformed into $\ln
\mathcal{M}$ (\ref{ln-M-call}).\smallskip
\end{enumerate}

It is here worth remarking that an analytical direct exponentiation of the
matrix $\mathcal{B}$ (\ref{dcoset2}), which would yield an explicit
symplectic frame \textit{e.g.} for the manifold (\ref{M-call-hat}) or (\ref
{M-call-tilde}), and which, through the 3-step procedure just mentioned,
would provide a more explicit form of the treatment of Secs. \ref{sec:56}
and \ref{manifest}, does not seem to be feasible (in \cite{BFSY-D=5} the
exponentiation of $\mathcal{B}$ (\ref{dcoset2}) with $t=0$ \textit{or},
equivalently, $t^{\prime }=0$ has only been performed). It may be possible
that a direct exponentiation of the matrix $\mathcal{B}$ (\ref{dcoset2})
could be performed by exploiting the fundamental identity for the $d$-tensor
of the symmetric coset. With the above normalization, such an identity can be
derived from the treatment given in \cite{Exc-Reds} (\textit{at least} for $%
E_{6}$):
\begin{equation}
d_{\alpha \beta \gamma }d^{\lambda \mu \gamma }d_{\mu \nu \rho }=d_{\nu \rho
\alpha }\delta _{\beta }^{\lambda }+\frac{1}{3}d_{\nu \rho \beta }\delta
_{\alpha }^{\lambda }-6d_{\mu \nu \rho }T_{I\mid \alpha }^{~~\lambda
}T_{~\beta }^{I~~\mu },  \label{ddd-E6-full}
\end{equation}
where, as in the explicit treatment of Secs. \ref{sec:56} and \ref{manifest}%
, Greek indices run over the fundamental $\mathbf{27}$ (or $\overline{%
\mathbf{27}}$) irrep., and capital Latin indices run over the adjoint $%
\mathbf{78}$ irrep. of $E_{6}$; the $T_{I\mid \alpha \beta }$'s denote the
realization of the generators of $E_{6}$ in its $\mathbf{27}$ irrep. (see
\textit{e.g.} \cite{Exc-Reds}), and they are \textit{e.g.} proportional to
the $\phi _{I}$'s (\ref{phi-split}) appearing in the matrices $Y_{I}$ (\ref
{Y_I}). We note that the complete symmetrization of covariant indices of the
identity (\ref{ddd-E6-full}) yields the well known identity:
\begin{equation}
d_{(\alpha \beta \mid \gamma }d^{\lambda \mu \gamma }d_{\mu \mid \nu \rho )}=%
\frac{4}{3}d_{(\nu \rho \beta }\delta _{\alpha )}^{\lambda }.
\end{equation}
We leave for the future the interesting task of exploiting the identity (\ref
{ddd-E6-full}) \textit{and/or} spectral techniques in order to perform the
exponentiation of the matrix $\mathcal{B}$ (\ref{dcoset2}), and thus to
determine a more explicit expression of the maximally manifestly covariant
symplectic frame introduced in Secs. \ref{sec:56} and \ref{manifest}%
.\bigskip\

\section{\label{Iwa-Exc}The Iwasawa Decomposition}

In this Section we are going to construct, along the lines of \cite{Iwa-N=8}%
, another parametrization for the coset $\mathcal{M}$ (\ref{M-call}), by
exploiting the \textit{Iwasawa decomposition}, which in this case turns out
to be manifestly $SO(8)$-covariant, thus providing a manifestly \textit{%
triality-symmetric} description of the rank-$3$ coset $\mathcal{M}$. Within
this treatment, we will denote by $\frak{P}$ the Lie algebra of the coset $%
\mathcal{M}$, namely the complement in $\frak{e}_{7\left( -25\right) }$ to
its maximal compact subalgebra $\frak{T}:=\frak{e}_{6\left( -78\right)
}\oplus \frak{u}\left( 1\right) $.\medskip

As the first step, one needs to determine a maximal non-compact Cartan
subalgebra $\frak{H}_{3}$. As observed at the start of Sec. \ref{manifest},
a possible choice is:
\begin{equation}
\frak{H}_{3}:=\langle Y_{123},Y_{132},Y_{133}\rangle _{\mathbb{R}}\subset
\frak{P},  \label{Cartan-M-call}
\end{equation}
generated by the diagonal elements of $\frak{J}_{3}(\mathbb{O})$.

Next, a basis of $54-3=51$ positive roots of $\frak{P}$ with respect to $%
\frak{H}_{3}$ (\ref{Cartan-M-call}) is to be determined.

If the \textit{adjoint action} of $\frak{H}_{3}$ on $\frak{e}_{7\left(
-25\right) }$ is simultaneously diagonalized, we expect to be able to find
only $102$ non-vanishing vectors in $\mathbb{R}^{3}$. This follows from the
fact that, apart from $\frak{H}_{3}$ itself, $\frak{H}_{3}$ commutes with a $%
28$-dimensional subalgebra $\frak{S}\simeq \frak{so}(8)$ of $\frak{T}$; this
can be easily understood by the following argument. By denoting with $\oplus
_{s}$ the semi-direct sum of algebras, due to the \textit{symmetric} nature
of the embedding determining the coset $\mathcal{M}$ (\ref{M-call}), the
structure of the Cartan decomposition of $\frak{e}_{7\left( -25\right) }=%
\frak{T}\oplus _{s}\frak{P}$ reads:
\begin{equation}
\lbrack \frak{T},\frak{T}]\subseteq \frak{T},\qquad \lbrack \frak{P},\frak{P}%
]\subseteq \frak{T},\qquad \lbrack \frak{T},\frak{P}]\subseteq \frak{P}.
\label{symm-Cartan}
\end{equation}
As usual, the last relation implies that $\frak{P}$ is a \textit{%
representation space} for $\frak{T}$, which in general will decompose in
\textit{irreducible} subspaces. In particular, $\frak{P}$ is a
representation space for the $\frak{f}_{4\left( -52\right) }$ subalgebra of $%
\frak{T}$. As it is well known, $\frak{f}_{4\left(-52\right) }$ is the Lie algebra
of the group Aut$\left( \frak{J}_{3}\left(
\mathbb{O}\right) \right) $; in turn the subalgebra of $\frak{f}_{4\left(
-52\right) }$ which keeps the diagonal elements of $\frak{J}_{3}\left(
\mathbb{O}\right) $ fixed is precisely $\frak{so}(8)$, namely the Lie
algebra of the automorphism group Aut$\left( \mathbf{t}\left( \mathbb{O}%
\right) \right) $ of the \textit{normed triality} on $\mathbb{O}$ (see
\textit{e.g.} \cite{Baez}). Therefore, since $\frak{H}_{3}$ has been
selected exactly as the subalgebra corresponding to the diagonal elements of
$\frak{J}_{3}\left( \mathbb{O}\right) $, it has to commute with an $\frak{so}%
(8)$ subalgebra of $\frak{f}_{4\left( -52\right) }$; this can indeed be
checked by inspecting the structure of the roots. Following \cite{F4,E6}, a
Cartan subalgebra $\frak{H}_{7}\subset \frak{e}_{7\left( -25\right) }$ can
be obtained by adding the space $\frak{H}_{4}$ generated by the four
matrices $Y_{i}$, $i=1,6,15,36$ (recall (\ref{Y_I})) to $\frak{H}_{3}$.
Then, the computation of the roots with respect to this system exactly
yields $28$ roots with vanishing components in the subspace $\frak{H}_{3}$;
these generate the $\frak{H}_{3}$-preserving Lie algebra \cite{F4}:
\begin{equation}
\frak{S}:=\left\langle Y_{1},\ldots Y_{21},Y_{30},\ldots
,Y_{36}\right\rangle _{\mathbb{R}}=\frak{so}\left( 8\right) ,
\label{S-fraktur}
\end{equation}
whose $\left\langle Y_{i}\right\rangle _{i=1,6,15,36}$ is thus a Cartan
subalgebra. Note that in \cite{F4,E6} a completion of $\left\langle
Y_{i}\right\rangle _{i=1,6,15,36}$ to $\frak{so}\left( 8\right) \neq \frak{S}
$ (\ref{S-fraktur}) was worked out, but this is irrelevant for the present
investigation.

As a consequence, it holds:
\begin{equation}
ad_{\frak{H}_{3}}|_{\frak{S}\oplus \frak{H}_{3}}=0,
\end{equation}
so that $31$ eigenvalues vanish in $\mathbb{R}^{3}$, and thus only at most $%
133-31=102$ can be non-vanishing, as expected.\smallskip

Let us show that actually all the remaining $102$ eigenvalues of $ad_{\frak{H%
}_{3}}$ on $\frak{e}_{7\left( -25\right) }$ are non-vanishing. First, we can
write:
\begin{eqnarray}
\frak{T} &=&\frak{S}\oplus _{s}\frak{T}^{\prime },  \label{T-decomp} \\
\frak{P} &=&\frak{H}_{3}\oplus _{s}\frak{P}^{\prime },  \label{P-decomp}
\end{eqnarray}
with dim$_{\mathbb{R}}\frak{T}^{\prime }=$dim$_{\mathbb{R}}\frak{P}^{\prime
}=51$. Now,
\begin{equation}
ad_{\frak{H}_{3}}:\frak{T}^{\prime }\oplus _{s}\frak{P}^{\prime
}\longrightarrow \frak{T}^{\prime }\oplus _{s}\frak{P}^{\prime }.
\end{equation}
Indeed, $[\frak{P},\frak{P}]\subseteq \frak{T}$ (\ref{symm-Cartan}) implies $%
[\frak{H}_{3},\frak{P}^{\prime }]\subseteq \frak{T}$. Let $\langle ,\rangle
_{ck}$ be the Cartan-Killing product; then, its restriction to $\frak{T}$
has a definite signature, usually chosen to be negative. The fact that $[%
\frak{S},\frak{H}_{3}]=0$ implies:
\begin{equation}
\langle \frak{S},[\frak{H}_{3},\frak{P}^{\prime }]\rangle _{ck}=-\langle
\lbrack \frak{H}_{3},\frak{S}],\frak{P}^{\prime }\rangle _{ck}=0\Rightarrow
\lbrack \frak{H}_{3},\frak{P}^{\prime }]\in \frak{T}^{\prime }.
\end{equation}
Next, from $[\frak{T},\frak{P}]\subseteq \frak{P}$ (\ref{symm-Cartan}) it
follows that $[\frak{H}_{3},\frak{T}^{\prime }]\subseteq \frak{P}$. As the
Cartan-Killing form is strictly positive on $\frak{P}$, and $\frak{H}_{3}$
is Abelian, one obtains:
\begin{equation}
\langle \frak{H}_{3},[\frak{H}_{3},\frak{T}^{\prime }]\rangle _{ck}=-\langle
\lbrack \frak{H}_{3},\frak{H}_{3}],\frak{T}^{\prime }\rangle
_{ck}=0\Rightarrow \lbrack \frak{H}_{3},\frak{T}^{\prime }]\subseteq \frak{P}%
^{\prime }.
\end{equation}
In this way, one can conclude that the set $\mathcal{W}$ of the remaining $%
102$ roots of $\frak{e}_{7\left( -25\right) }$ has eigenspaces in $\frak{T}%
^{\prime }\oplus _{s}\frak{P}^{\prime }$. Thus, each eigenvector has the
form $\lambda _{A}:=t_{A}+p_{A}$, $A=1,...,102$, where $t_{A}\in \frak{T}%
^{\prime }$ and $p_{A}\in \frak{P}^{\prime }$ are both non-vanishing and
uniquely determined by $\lambda _{A}$. Let us suppose that one of the roots $%
r_{A_{0}}\in \mathcal{W}$ vanishes: $r_{A_{0}}=0$. This would imply that $%
ad_{\frak{H}_{3}}(p_{A_{0}})=0$. But, in turn, this would also mean $%
p_{A_{0}}\in \frak{H}_{3}$ (as $\frak{H}_{3}$ is a \textit{maximal} Cartan
subalgebra in $\frak{P}$), which cannot be the case, since $\frak{H}_{3}\cap
\frak{P}^{\prime }=0 $.\newline
Keeping in mind that we are considering the roots of $\frak{e}_{7\left(
-25\right) }$ relative to the choice (\ref{Cartan-M-call}) of $\frak{H}_{3}$%
, we can thus conclude that all $102$ roots in $\mathcal{W}$ are
non-vanishing. $\blacksquare \smallskip $

Let us now fix a choice of $51$ positive roots $\mathcal{W}_{+}$ so that $%
\mathcal{W}=\mathcal{W}_{+}\cup \mathcal{W}_{-}$. The corresponding
eigenspaces are one-dimensional, and they are generated by the eigenvectors $%
\lambda _{i}^{+}$, $i=1,\ldots ,51$, with eigenvalues $r_{i}\in \mathcal{W}%
_{+}$. We can write in a unique way:
\begin{equation}
\lambda _{i}^{+}=p_{i}+t_{i},  \label{pos-eigen}
\end{equation}
{implying} that:
\begin{equation}
\lambda _{i}^{-}:=p_{i}-t_{i}
\end{equation}
are eigenvalues of $-r_{i}\in \mathcal{W}_{-}$.\newline

Finally, by renaming $h_{1}:=Y_{123}$, $h_{2}:=Y_{132}$ and $h_{3}:=Y_{133}$%
, the \textit{Iwasawa decomposition} of the coset $\mathcal{M}$ (\ref{M-call}%
) can be written as:
\begin{equation}
\mathcal{M}:=\exp (x_{1}h_{1}+x_{2}h_{2}+x_{3}h_{3})\exp
(\sum_{i=1}^{51}y_{i}\lambda _{i}^{+}).  \label{Iwa-parametrization}
\end{equation}
which exhibits a manifest $SO(8)$-covariance. We anticipate that $\frak{so}%
\left( 8\right) $ is the Lie algebra of Aut$\left( \mathbf{t}\left( \mathbb{O%
}\right) \right) =Spin\left( 8\right) $, namely the automorphism group of
the \textit{normed triality} $\mathbf{t}\left( \mathbb{O}\right) $ over the
division algebra of octonions $\mathbb{O}$ (see \textit{e.g.} \cite{Baez}):
\begin{equation}
\frak{so}\left( 8\right) =\frak{Aut}\left( \mathbf{t}\left( \mathbb{O}%
\right) \right) =:\frak{tri}\left( \mathbb{O}\right) ;  \label{so(8)-anticip}
\end{equation}
see the discussion in Sec. \ref{Generalizations}.


\subsection{\label{Manifest-Covariance}$SO\left( 8\right) $-\textit{Triality}%
}

Now, we want study the $SO(8)$-covariance of the Iwasawa parametrization (%
\ref{Iwa-parametrization}) in more detail.

First, as pointed out above, the elements $h_{1}$, $h_{2}$, $h_{3}$ of the
Cartan subalgebra $\frak{H}_{3}$ commute with $SO(8)$, and it follows that
they are three $SO(8)$-singlets. Thus, the $51$-dimensional linear space $%
\Lambda _{+}$ generated by the positive roots $\mathcal{W}_{+}$ is invariant
under the (adjoint) action of $SO(8)$, and it decomposes into irreps. of $%
SO(8)$ as:
\begin{equation}
\Lambda _{+}=\mathbf{1}^{3}+\mathbf{8}_{v}^{2}+\mathbf{8}_{c}^{2}+\mathbf{8}%
_{s}^{2}.  \label{triality}
\end{equation}

The manifestly \textit{triality-symmetric} decomposition (\ref{triality})
can be proven by means of the following general argument. Let us fix an
orthonormal basis $L_{1},\ldots ,L_{7}$ of $\mathbb{R}^{7}$. Then, the $%
\left( 133-7\right) /2=63$ positive roots of $E_{7\left( -25\right) }$ can
be represented as (see \textit{e.g.} \cite{fultonharris}, p. 333):
\begin{eqnarray}
&&L_{m}\pm L_{n},\qquad 1\leq n<m\leq 6;\qquad \sqrt{2}L_{7};  \notag \\
&&\frac{\pm L_{1}\pm L_{2}\pm L_{3}\pm L_{4}\pm L_{5}\pm L_{6}+\sqrt{2}L_{7}%
}{2},\quad \text{odd~number~of~}-~\text{signs}.
\end{eqnarray}
Among these, the $\left( 28-4\right) /2=12$ roots:
\begin{equation}
\mu _{mn}^{\pm }:=L_{m}\pm L_{n},\qquad 1\leq n<m\leq 4  \label{mu}
\end{equation}
are the positive roots of $\frak{so}(8)$. $\mu _{mn}^{\pm }$ (\ref{mu})
provide a representation of the algebra $\frak{so}(8)$ over the linear space
generated by the remaining $51$ roots in the usual way: if, consistent with (%
\ref{pos-eigen}), we call $\lambda _{i}^{+}$ the $51$ complementary roots,
then the linear operators $\mu _{mn}^{\pm }$ and their corresponding \textit{%
adjoint} $\tilde{\mu}_{mn}^{\pm }$ are defined by:
\begin{eqnarray}
&&\mu _{mn}^{\pm }(\lambda _{i}^{+}):=\lambda _{i}^{+}+(L_{m}\pm L_{n}), \\
&&\tilde{\mu}_{mn}^{\pm }(\lambda _{i}^{+}):=\lambda _{i}^{+}-(L_{m}\pm
L_{n}),
\end{eqnarray}
where the result is intended to be zero when the vectors on the right-hand
side are not roots.\newline
\smallskip

{T}his procedure allows to identify exactly $9$ invariant subspaces of $%
\Lambda _{+}$:

\begin{enumerate}
\item  The three spaces respectively generated by $L_{6}+L_{5}$, $%
L_{6}-L_{5} $ and $\sqrt{2}L_{7}$ are one-dimensional invariant subspaces
defining a $\mathbf{1}^{3}$ representation (sum of $3$ $SO\left( 8\right) $%
-singlets).

\item  The two $8$-dimensional spaces $V_{5}$ and $V_{6}$ respectively
generated by the basis:
\begin{eqnarray}
&&\{L_{5}\pm L_{n}\}_{n=1}^{4}; \\
&&\{L_{6}\pm L_{n}\}_{n=1}^{4};
\end{eqnarray}
are both representations with weights $\pm L_{n}$, and thus correspond to
two copies of the \textit{vector} representation $\mathbf{8}_{v}$.

\item  The two $8$-dimensional spaces $C_{+}$ and $C_{-}$ respectively
generated by the basis:
\begin{eqnarray}
&&\frac{\pm L_{1}\pm L_{2}\pm L_{3}\pm L_{4}+L_{5}-L_{6}+\sqrt{2}L_{7}}{2};
\\
&&\frac{\pm L_{1}\pm L_{2}\pm L_{3}\pm L_{4}-L_{5}+L_{6}+\sqrt{2}L_{7}}{2};
\end{eqnarray}
(with an \textit{even} number of $-$ signs) are both representations with
weights $\frac{\pm L_{1}\pm L_{2}\pm L_{3}\pm L_{4}}{2}$ (with an \textit{%
even} number of $-$ signs), thus providing two copies of the \textit{chiral
spinor} representation $\mathbf{8}_{c}$.

\item  The two $8$-dimensional spaces $S_{+}$ and $S_{-}$ respectively
generated by the basis:
\begin{eqnarray}
&&\frac{\pm L_{1}\pm L_{2}\pm L_{3}\pm L_{4}+L_{5}+L_{6}+\sqrt{2}L_{7}}{2};
\\
&&\newline
\frac{\pm L_{1}\pm L_{2}\pm L_{3}\pm L_{4}-L_{5}-L_{6}+\sqrt{2}L_{7}}{2},
\end{eqnarray}
(with an \textit{odd} number of $-$ signs) are both representations with
weights $\frac{\pm L_{1}\pm L_{2}\pm L_{3}\pm L_{4}}{2}$ (with an \textit{odd%
} number of $-$ signs), thus providing two copies of the \textit{chiral
spinor} representation $\mathbf{8}_{s}$ (conjugate of $\mathbf{8}_{c}$).
\end{enumerate}

This implies (\ref{triality}), in which the $SO(8)$\textit{-triality} is
manifest. It is worth remarking that the appearance of the square for the
three $\mathbf{8}$ irreps. in (\ref{triality}) is a consequence of the
complex (in particular, \textit{special K\"{a}hler}, as mentioned in
previous Sections) structure of the coset $\mathcal{M}$ (\ref{M-call}).


\subsection{\label{Group-Theory}Group Theory}

In the \textit{Iwasawa parametrization} of $\mathcal{M}$ (\ref{M-call})
worked out in Sec. \ref{Iwa-Exc}, the resulting maximal manifest covariance
group is nothing but the $SO\left( 8\right) $ group (uniquely determined in $%
E_{7\left( -25\right) }$; see Subsubsecs. \ref{First} and \ref{Second})
preserving the diagonal elements in the rank-$3$ simple Jordan algebra $%
\frak{J}_{3}\left( \mathbb{O}\right) $. As clearly evident form the chain (%
\ref{1-bis}) of embeddings, such an $SO\left( 8\right) $ is placed as
follows:
\begin{equation}
SO\left( 8\right) \subset \left[ (SO(10)\times U(1))\cap F_{4(-52)}\right] .
\label{SO(8)-def}
\end{equation}
As given by (\ref{so(8)-anticip}), it shares the same algebra $\frak{so}(8)=%
\frak{tri}(\mathbb{O})$ with the automorphism group Aut$\left( \mathbf{t}%
\left( \mathbb{O}\right) \right) =Spin\left( 8\right) $ of the \textit{%
normed triality} over the octonions $\mathbb{O}$ \cite{Baez}. Furthermore,
it is worth remarking that such an $SO\left( 8\right) $ recently appeared as
the stabilizer of the BPS generic charge orbit in the two-centered extremal
black hole solutions of $\mathcal{N}=2$, $D=4$ exceptional supergravity; see Table \ref{BPS} \cite{Irred-1}.

We also note that, at the level of (manifest) covariance, the Iwasawa
parametrization of $\mathcal{M}$ (\ref{M-call}) worked out in Sec. \ref
{Iwa-Exc} differs from the Iwasawa parametrization of $\mathcal{M}_{\mathcal{%
N}=8}$ (\ref{M-call-N=8}) studied in \cite{Iwa-N=8}, whose manifest maximal
(non-compact) covariance is $SL\left( 7,\mathbb{R}\right) $, with maximal
compact subgroup $SO\left( 7\right) $.

\subsubsection{\label{First}A First Chain of Embeddings}

The chain of maximal symmetric embeddings relevant for the study of the
maximal manifest covariance of the Iwasawa parametrization (\ref
{Iwa-parametrization}) of the irreducible Riemannian globally symmetric rank-%
$3$ symmetric special K\"{a}hler coset $\mathcal{M}$ (\ref{M-call}) reads as
follows (see \textit{e.g.} \cite{Gilmore}):

\begin{enumerate}
\item  in the \textit{compact} case:
\begin{eqnarray}
E_{7\left( -25\right) } &\supset &E_{6\left( -78\right) }\times U\left(
1\right) ^{\prime }  \notag \\
&\supset &SO\left( 10\right) \times U\left( 1\right) ^{\prime }\times
U\left( 1\right) ^{\prime \prime }  \notag \\
&\supset &SO\left( 8\right) \times U(1)^{\prime }\times U(1)^{\prime \prime
}\times U(1)^{\prime \prime \prime };  \label{1-bis}
\end{eqnarray}

\item  in the relevant (namely, \textit{minimally}) \textit{non-compact}
case:
\begin{eqnarray}
E_{7\left( -25\right) } &\supset &E_{6\left( -26\right) }\times SO\left(
1,1\right) ^{\prime }  \notag \\
&\supset &SO\left( 9,1\right) \times SO\left( 1,1\right) ^{\prime }\times
SO\left( 1,1\right) ^{\prime \prime }  \notag \\
&\supset &SO\left( 8\right) \times SO\left( 1,1\right) ^{\prime }\times
SO\left( 1,1\right) ^{\prime \prime }\times SO(1,1)^{\prime \prime \prime }.
\label{2-bis}
\end{eqnarray}
In the last line of (\ref{2-bis}) the first two $SO\left( 1,1\right) $
factors have the physical meaning of ``extra'' $T$-dualities generated by
the Kaluza-Klein reductions, respectively $D=5\rightarrow D=4$, and $%
D=6\rightarrow D=5$.
\end{enumerate}

Correspondingly, the adjoint irrep. $\mathbf{133}$ of $E_{7\left( -25\right)
}$ branches as (subscripts denote $U\left( 1\right) $-charges or $SO\left(
1,1\right) $-weights, for (\ref{1-bis}) and (\ref{2-bis}) respectively,
throughout; see \textit{e.g.} \cite{Slansky}):
\begin{eqnarray}
\mathbf{133} &=&\mathbf{78}_{0}+\mathbf{1}_{0}+\mathbf{27}_{-2}+\mathbf{27}%
_{+2}^{\prime }  \label{3-bis} \\
&&  \notag \\
&=&\mathbf{1}_{0,0}+\mathbf{16}_{0,-3}+\mathbf{16}_{0,+3}^{\prime }+\mathbf{%
45}_{0,0}+\mathbf{1}_{0,0}  \notag \\
&&+\mathbf{1}_{-2,+4}+\mathbf{10}_{-2,-2}+\mathbf{16}_{-2,+1}  \notag \\
&&+\mathbf{1}_{+2,-4}+\mathbf{10}_{+2,+2}+\mathbf{16}_{+2,-1}^{\prime }
\notag \\
&&  \notag \\
&=&\mathbf{1}_{0,0,0}+\mathbf{8}_{c,0,-3,1}+\mathbf{8}_{s,0,-3,-1}+\mathbf{8}%
_{c,0,+3,-1}+\mathbf{8}_{s,0,+3,+1}  \label{4-bis} \\
&&+\mathbf{1}_{0,0,0}+\mathbf{8}_{v,0,0,+2}+\mathbf{8}_{v,0,0,-2}+\mathbf{28}%
_{0,0,0}+\mathbf{1}_{0,0,0}  \label{5-bis} \\
&&+\mathbf{1}_{-2,+4,0}+\mathbf{1}_{-2,-2,+2}+\mathbf{1}_{-2,-2,-2}+\mathbf{8%
}_{v,-2,-2,0}+\mathbf{8}_{c,-2,+1,+1}+\mathbf{8}_{s,-2,+1,-1}  \label{6} \\
&&+\mathbf{1}_{+2,-4,0}+\mathbf{1}_{+2,+2,-2}+\mathbf{1}_{+2,+2,+2}+\mathbf{8%
}_{v,+2,+2,0}+\mathbf{8}_{c,+2,-1,-1}+\mathbf{8}_{s,+2,-1,+1}.  \label{7}
\end{eqnarray}
Recalling the treatment Sec. \ref{Iwa-Exc}, in line (\ref{3-bis}) one can
recognize:
\begin{eqnarray}
\frak{T} &:&=\mathbf{Adj}\left( E_{6}\times U\left( 1\right) \right) =%
\mathbf{78}_{0}+\mathbf{1}_{0};  \label{T-def} \\
\frak{P} &:&=\mathbf{27}_{-2}+\mathbf{27}_{+2}^{\prime },  \label{P-def}
\end{eqnarray}
where, by the definitions introduced in\ Sec. \ref{Iwa-Exc}, $\frak{P}$
denotes (the irreducible decomposition of) the Lie algebra of the coset $%
\mathcal{M}$ (\ref{M-call}) (as representation space of $\frak{T}$).
Furthermore, $\mathbf{27}_{-2}+\mathbf{27}_{+2}^{\prime }$ manifestly shows
the complex (\textit{special K\"{a}hler}) structure of $\mathcal{M}$ itself,
which is then spoiled by the further subsequent branchings needed for the
Iwasawa parametrization (\ref{Iwa-parametrization}).

Furthermore, the lines (\ref{4-bis}) and (\ref{5-bis}) give the $SO\left(
8\right) \times \left[ U\left( 1\right) \right] ^{3}$ (or $SO\left( 8\right)
\times \left[ SO\left( 1,1\right) \right] ^{3}$) irreducible branching of
the $79$ compact generators of $E_{7\left( -25\right) }$, namely of the
generators of its maximal compact subgroup $E_{6\left( -78\right) }\times
U\left( 1\right) $. On the other hand, the lines (\ref{6}) and (\ref{7})
give the $SO\left( 8\right) \times \left[ U\left( 1\right) \right] ^{3}$ (or
$SO\left( 8\right) \times \left[ SO\left( 1,1\right) \right] ^{3}$)
irreducible branching of the $54$ non-compact generators of $E_{7\left(
-25\right) }$, namely of the generators of $\mathcal{M}$ itself. In
particular, recalling the definitions of Sec. \ref{Iwa-Exc}:
\begin{eqnarray}
\frak{T}^{\prime } &:&=\mathbf{1}_{0,0,0}+\mathbf{8}_{c,0,-3,1}+\mathbf{8}%
_{s,0,-3,-1}+\mathbf{8}_{c,0,+3,-1}+\mathbf{8}_{s,0,+3,+1}  \notag \\
&&+\mathbf{1}_{0,0,0}+\mathbf{8}_{v,0,0,+2}+\mathbf{8}_{v,0,0,-2}+\mathbf{1}%
_{0,0,0}; \\
\frak{S} &:&=\mathbf{Adj}\left( SO\left( 8\right) \right) =\mathbf{28}%
_{0,0,0}.
\end{eqnarray}
$\frak{T}^{\prime }$ is the Lie algebra of the \textit{non-maximal} (and
\textit{non-symmetric}) coset (dim$_{\mathbb{R}}=51$):
\begin{equation}
\frac{E_{6\left( -78\right) }\times U\left( 1\right) ^{\prime }}{SO\left(
8\right) }=\frac{E_{6\left( -78\right) }}{SO\left( 8\right) }\times U\left(
1\right) ^{\prime },
\end{equation}
or, in the choice of chain (\ref{2-bis}), of its relevant (\textit{i.e.
minimally}) non-compact form:
\begin{equation}
\frac{E_{6\left( -26\right) }\times SO\left( 1,1\right) ^{\prime }}{SO\left(
8\right) }=\frac{E_{6\left( -26\right) }}{SO\left( 8\right) }\times SO\left(
1,1\right) ^{\prime }.
\end{equation}
Out of the six $SO\left( 8\right) $-singlets:
\begin{equation}
\mathbf{1}_{-2,+4,0},~\mathbf{1}_{-2,-2,+2},~\mathbf{1}_{-2,-2,-2},~\mathbf{1%
}_{+2,-4,0},~\mathbf{1}_{+2,+2,-2},~\mathbf{1}_{+2,+2,+2}
\label{six-SO(8)-singlets}
\end{equation}
in lines (\ref{6}) and (\ref{7}), three linear combinations generate $\frak{H%
}_{3}$, whereas the remaining linear combinations, orthogonal with respect
to the Cartan-Killing form, together with the manifestly $SO\left( 8\right) $%
\textit{-triality-symmetric} branching:
\begin{eqnarray}
&&\mathbf{8}_{v,-2,-2,0}+\mathbf{8}_{c,-2,+1,+1}+\mathbf{8}_{s,-2,+1,-1}
\notag \\
&&+\mathbf{8}_{v,+2,+2,0}+\mathbf{8}_{c,+2,-1,-1}+\mathbf{8}_{s,+2,-1,+1}
\end{eqnarray}
of lines (\ref{6}) and (\ref{7}), generate $\frak{P}^{\prime }$.

Analogously, the smallest non-trivial symplectic irrep., namely the
fundamental $\mathbf{56}$ of $E_{7\left( -25\right) }$, branches as (see
\textit{e.g.} \cite{Slansky}):
\begin{eqnarray}
\mathbf{56} &=&\mathbf{27}_{+1}+\mathbf{27}_{-1}^{\prime }+\mathbf{1}_{+3}+%
\mathbf{1}_{-3}  \label{56-branching-1} \\
&&  \notag \\
&=&\mathbf{1}_{+1,+4}+\mathbf{10}_{+1,-2}+\mathbf{16}_{+1,+1}  \notag \\
&&+\mathbf{1}_{-1,-4}+\mathbf{10}_{-1,+2}+\mathbf{16}_{-1,-1}^{\prime }
\notag \\
&&+\mathbf{1}_{+3,0}+\mathbf{1}_{-3,0}  \label{56-branching-2} \\
&&  \notag \\
&=&\mathbf{1}_{+1,+4,0}+\mathbf{1}_{+1,-2,+2}+\mathbf{1}_{+1,-2,-2}+\mathbf{8%
}_{v,+1,-2,0}+\mathbf{8}_{c,+1,+1,+1}+\mathbf{8}_{s,+1,+1,-1}  \notag \\
&&+\mathbf{1}_{-1,-4,0}+\mathbf{1}_{-1,+2,-2}+\mathbf{1}_{-1,+2,+2}+\mathbf{8%
}_{v,-1,+2,0}+\mathbf{8}_{c,-1,-1,-1}+\mathbf{8}_{s,-1,-1,+1}  \notag \\
&&+\mathbf{1}_{+3,0,0}+\mathbf{1}_{-3,0,0},  \label{56-branching-3}
\end{eqnarray}

It is instructive to analyze the branchings (\ref{56-branching-1})-(\ref
{56-branching-3}) more in depth.

{From} the structure of the matrices $Y_{I}$ (\ref{Y_I}) and (\ref{Y_79}), $%
I=1,\ldots ,79$, the structure of the first branching (\ref{56-branching-1})
is evident, where the subscripts denote the charge (weight) with respect to $%
U\left( 1\right) ^{\prime }$ ($SO\left( 1,1\right) ^{\prime }$). Next, let
us look at the $\mathbf{27}$ irrep. of $E_{6}$; which is realized over the $%
27$ dimensional linear space of octonionic Hermitian matrices underlying the
exceptional Jordan algebra $\frak{J}_{3}\left( \mathbb{O}\right) $. It
decomposes as follows:
\begin{equation}
\underset{\mathbf{27}}{
\begin{pmatrix}
a & X & Y \\
X^{\ast } & b & Z \\
Y^{\ast } & Z^{\ast } & c
\end{pmatrix}
}=\underset{\mathbf{16}}{
\begin{pmatrix}
0 & X & Y \\
X^{\ast } & 0 & 0 \\
Y^{\ast } & 0 & 0
\end{pmatrix}
}+\underset{\mathbf{10}}{
\begin{pmatrix}
0 & 0 & 0 \\
0 & b & Z \\
0 & Z^{\ast } & c
\end{pmatrix}
}+\underset{\mathbf{1}}{
\begin{pmatrix}
a & 0 & 0 \\
0 & 0 & 0 \\
0 & 0 & 0
\end{pmatrix}
},  \label{27-decomp}
\end{equation}
where $a,b,c$ are real numbers and $X,Y,Z$ are real (in the linear sense)
octonions. As hinted in (\ref{27-decomp}), this yields a decomposition $%
\mathbf{27}=\mathbf{16}+\mathbf{10}+\mathbf{1}$ of invariant spaces under
the maximal symmetric subgroup $R:=SO(10)\times U(1)^{\prime \prime }$ of $%
E_{6}$ (we consider, without loss of any generality, the compact chain (\ref
{1-bis}) of embeddings). Indeed, the one-dimensional space $\mathbf{1}$ is
easily seen to be invariant under $R$. As the spaces in the decomposition (%
\ref{27-decomp}) are orthogonal with respect to the trace product (which is
preserved by $R$), its complement is also $R$-invariant. On the other hand,
the $16$-dimensional subspace defines the largest subalgebra in $\frak{J}%
_{3}\left( \mathbb{O}\right) $ complementary to the one-dimensional space $%
\mathbf{1}$. This proves our assertion.

The $U(1)^{\prime \prime }$-charges of the spaces in the right-hand side of (%
\ref{27-decomp}) can be determined by noting that $U(1)^{\prime \prime
}\nsubseteq F_{4\left( -52\right) }$. From the treatment of Sec. \ref{sec:56}%
, the Lie algebra $\frak{e}_{6\left( -78\right) }$ is obtained by adding the
left (or right) action of $\frak{J}_{3}^{\prime }(\mathbb{O})$ on $\frak{J}%
_{3}(\mathbb{O})$ (where the prime here denotes the matrix tracelessness).
This means that the generator of $U(1)^{\prime \prime }$ must be realized by
a traceless matrix $C_{U(1)^{\prime \prime }}$ in $\frak{J}_{3}(\mathbb{O})$
that by left Jordan-multiplication acts proportionally to the identity on
the three subspaces of the decomposition (\ref{27-decomp}). This implies
that:
\begin{equation}
C_{U\left( 1\right) ^{\prime \prime }}=
\begin{pmatrix}
2\gamma & 0 & 0 \\
0 & -\gamma & 0 \\
0 & 0 & -\gamma
\end{pmatrix}
.
\end{equation}
Writing (\ref{27-decomp}) as $V_{27}=V_{16}+V_{10}+V_{1}$, we see that:
\begin{equation}
C_{U\left( 1\right) ^{\prime \prime }}\circ V_{16}=\frac{\gamma }{2}%
V_{16},\qquad C_{U\left( 1\right) ^{\prime \prime }}\circ V_{10}=-\gamma
V_{10},\qquad \ C_{U\left( 1\right) ^{\prime \prime }}\circ V_{1}=2\gamma
V_{1}.
\end{equation}
By choosing the normalization of the charges in such the way that $\exp
(xC_{U\left( 1\right) ^{\prime \prime }})$ has period $2\pi $, one then
obtains:
\begin{equation}
\mathbf{27}=\mathbf{16}_{1}+\mathbf{10}_{-2}+\mathbf{1}_{4},
\end{equation}
which matches the convention \textit{e.g.} of \cite{Slansky}. Obviously, $%
U(1)^{\prime \prime }$ commutes with $U(1)^{\prime }$; therefore (\ref
{56-branching-2}) is obtained.

For the last branching (\ref{56-branching-3}), the decompositions of $%
\mathbf{10}$ and $\mathbf{16}$ have to be analyzed. As $SO(8)$ leaves the
diagonal matrices of $\frak{J}_{3}(\mathbb{O})$ invariant, it follows that
under its action the space $V_{10}$ decomposes as $%
V_{10}=V_{8}+V_{1,I}+V_{1,II}$ in the following way:
\begin{equation}
\underset{\mathbf{10}}{
\begin{pmatrix}
0 & 0 & 0 \\
0 & b & Z \\
0 & Z^{\ast } & c
\end{pmatrix}
}=\underset{\mathbf{8}_{v}}{
\begin{pmatrix}
0 & 0 & 0 \\
0 & 0 & Z \\
0 & Z^{\ast } & 0
\end{pmatrix}
}+\underset{\mathbf{1}_{I}}{
\begin{pmatrix}
0 & 0 & 0 \\
0 & b & 0 \\
0 & 0 & 0
\end{pmatrix}
}+\underset{\mathbf{1}_{II}}{
\begin{pmatrix}
0 & 0 & 0 \\
0 & 0 & 0 \\
0 & 0 & c
\end{pmatrix}
}.
\end{equation}
In order to determine the $U(1)^{\prime \prime \prime }$ charges, we again
observe that $U(1)^{\prime \prime \prime }\varsubsetneq F_{4\left(
-52\right) }$. Moreover, the $U(1)^{\prime \prime \prime }$ charge of $%
\mathbf{1}_{1,4}$ in (\ref{56-branching-2}) must be zero and, therefore, the
$U(1)^{\prime \prime \prime }$ generator must be realized by a matrix of the
form:
\begin{equation}
C_{U(1)^{\prime \prime \prime }}=
\begin{pmatrix}
0 & 0 & 0 \\
0 & \gamma ^{\prime } & 0 \\
0 & 0 & -\gamma ^{\prime }
\end{pmatrix}
.  \label{'''}
\end{equation}
By choosing the normalization as before in such a way that the period of $%
\exp (xC_{U(1)^{\prime \prime \prime }})$ is $2\pi $, one can fix $\gamma
^{\prime }=2$, and the charges turn out to be $0$, $2$ and $-2$ for $%
V_{8},\,V_{1,I}$ and $V_{1,II}$ respectively. Since $V_{8}$ is contained in
the vector representation $V_{10}$ of $SO(10)$, it has to correspond to the
vector rep. $\mathbf{8}_{v}$ of $SO\left( 8\right) $, so that:
\begin{equation}
\mathbf{10}=\mathbf{8}_{v,0}+\mathbf{1}_{2}+\mathbf{1}_{-2}.
\end{equation}
The charge operator $C_{U(1)^{\prime \prime \prime }}$ (\ref{'''}) splits $%
V_{16}$ into eigenspaces $V_{8}^{+}$ and $V_{8}^{-}$ with eigenvalues $1$
and $-1$, respectively:
\begin{equation}
\underset{\mathbf{16}}{
\begin{pmatrix}
0 & X & Y \\
X^{\ast } & 0 & 0 \\
Y^{\ast } & 0 & 0
\end{pmatrix}
}=\underset{\mathbf{8}_{c}}{
\begin{pmatrix}
0 & X & 0 \\
X^{\ast } & 0 & 0 \\
0 & 0 & 0
\end{pmatrix}
}+\underset{\mathbf{8}_{s}}{
\begin{pmatrix}
0 & 0 & Y \\
0 & 0 & 0 \\
Y^{\ast } & 0 & 0
\end{pmatrix}
}.
\end{equation}
The weights of $\mathbf{16}$ are $\frac{1}{2}\{\epsilon _{1},\ldots
,\epsilon _{8}\}$ where the $\epsilon $'s can assume all possible signs;
this means that $\mathbf{16}$ breaks into the direct sum of the conjugate
irreducible spinor representations $\mathbf{8}_{s}$ and $\mathbf{8}_{c}$ of $%
SO(8)$, with $\frac{1}{2}C_{U(1)^{\prime \prime \prime }}$ measuring their
chirality:
\begin{equation}
\mathbf{16}=\mathbf{8}_{s,-1}+\mathbf{8}_{c,1},
\end{equation}
which allows one to recover (\ref{56-branching-3}).

\subsubsection{\label{Second}A Second Chain of Embeddings}

A second chain of maximal and symmetric embeddings, relevant in order to
highlight the relation to the symmetry groups of the rank-$3$ Euclidean
Jordan algebra $\frak{J}_{3}\left( \mathbb{O}\right) $ and also for a
subsequent generalization at least for all \textit{conformal non-compact}
form of \textit{non-degenerate} \cite{Garibaldi} groups of type $E_{7}$ \cite
{Brown} (see Sec. \ref{Generalizations}), reads as follows:
\begin{eqnarray}
E_{7\left( -25\right) } &\supset &E_{6\left( -26\right) }\times SO\left(
1,1\right) ^{\prime }  \notag \\
&\supset &F_{4\left( -52\right) }\times SO\left( 1,1\right) ^{\prime }
\notag \\
&\supset &SO\left( 9\right) \times SO\left( 1,1\right) ^{\prime }  \notag \\
&\supset &SO\left( 8\right) \times SO\left( 1,1\right) ^{\prime },
\label{second}
\end{eqnarray}
where $SO\left( 8\right) $ in the fourth line of (\ref{second}) coincides
with the $SO\left( 8\right) $ in the third line of (\ref{1-bis}) and (\ref
{2-bis}). Moreover, (\ref{second}) also clarifies (\ref{SO(8)-def}). As already mentioned above, it holds that (see \textit{e.g.} \cite{G-Lects,Small-Orbits-Phys,Small-Orbits-Maths}):
\begin{eqnarray}
E_{7\left( -25\right) } &=&\text{Conf}\left( \frak{J}_{3}\left( \mathbb{O}%
\right) \right) =\text{Aut}\left[ \frak{M}\left( \frak{J}_{3}\left( \mathbb{O%
}\right) \right) \right] =G_{4};  \label{deff-1} \\
E_{6\left( -26\right) } &=&\text{Str}_{0}\left( \frak{J}_{3}\left( \mathbb{O}%
\right) \right) =G_{5};  \label{deff-2} \\
F_{4\left( -52\right) } &=&\text{Aut}\left( \frak{J}_{3}\left( \mathbb{O}%
\right) \right) =\text{mcs}\left[ \text{Str}_{0}\left( \frak{J}_{3}\left(
\mathbb{O}\right) \right) \right] ;  \label{deff-3} \\
\frak{so}\left( 8\right) &=&\frak{Aut}\left( \mathbf{t}\left( \mathbb{O}%
\right) \right) =:\frak{tri}(O),  \label{deff-4}
\end{eqnarray}
where $\mathbf{t}\left( \mathbb{O}\right) $ denotes the \textit{normed
triality} over the octonions $\mathbb{O}$ (see \textit{e.g.} \cite{Baez}), and (\ref{so(8)-anticip}) has been recalled. In (\ref
{second}), $SO\left( 1,1\right) $ has the physical meaning of ``extra'' $T$%
-duality generated by the Kaluza-Klein reduction $D=5\rightarrow D=4$.

Correspondingly, the adjoint irrep. $\mathbf{133}$ of $E_{7\left( -25\right)
}$ branches as (see \textit{e.g.} \cite{Slansky}):
\begin{eqnarray}
\mathbf{133} &\rightarrow &\mathbf{78}_{0}+\mathbf{1}_{0}+\mathbf{27}_{-2}+%
\mathbf{27}_{+2}^{\prime }  \notag \\
&&  \notag \\
&\rightarrow &\mathbf{26}_{0}+\mathbf{52}_{0}+\mathbf{1}_{0}+\mathbf{1}_{-2}+%
\mathbf{26}_{-2}+\mathbf{1}_{+2}+\mathbf{26}_{+2}  \notag \\
&&  \notag \\
&\rightarrow &\mathbf{1}_{0}+\mathbf{9}_{0}+\mathbf{16}_{0}+\mathbf{16}_{0}+%
\mathbf{36}_{0}+\mathbf{1}_{0}  \notag \\
&&+\mathbf{1}_{-2}+\mathbf{1}_{-2}+\mathbf{9}_{-2}+\mathbf{16}_{-2}  \notag
\\
&&+\mathbf{1}_{+2}+\mathbf{1}_{+2}+\mathbf{9}_{+2}+\mathbf{16}_{+2}  \notag
\\
&&  \notag \\
&\rightarrow &\mathbf{1}_{0}+\mathbf{1}_{0}+\mathbf{8}_{v,0}+\mathbf{8}%
_{c,0}+\mathbf{8}_{s,0}+\mathbf{8}_{c,0}+\mathbf{8}_{s,0}+\mathbf{8}_{v,0}+%
\mathbf{28}_{0}+\mathbf{1}_{0}  \notag \\
&&+\mathbf{1}_{-2}+\mathbf{1}_{-2}+\mathbf{1}_{-2}+\mathbf{8}_{v,-2}+\mathbf{%
8}_{c,-2}+\mathbf{8}_{s,-2}  \notag \\
&&+\mathbf{1}_{+2}+\mathbf{1}_{+2}+\mathbf{1}_{+2}+\mathbf{8}_{v,+2}+\mathbf{%
8}_{c,+2}+\mathbf{8}_{s,+2}.
\end{eqnarray}

Analogously, the symplectic fundamental irrep. $\mathbf{56}$ of $E_{7\left(
-25\right) }$ branches as (see \textit{e.g.} \cite{Slansky}):
\begin{eqnarray}
\mathbf{56} &\rightarrow &\mathbf{27}_{+1}+\mathbf{27}_{-1}^{\prime }+%
\mathbf{1}_{+3}+\mathbf{1}_{-3}  \notag \\
&&  \notag \\
&\rightarrow &\mathbf{1}_{+1}+\mathbf{26}_{+1}+\mathbf{1}_{-1}+\mathbf{26}%
_{-1}+\mathbf{1}_{+3}+\mathbf{1}_{-3}  \notag \\
&&  \notag \\
&\rightarrow &\mathbf{1}_{+1}+\mathbf{1}_{+1}+\mathbf{9}_{+1}+\mathbf{16}%
_{+1}+\mathbf{1}_{-1}+\mathbf{1}_{-1}+\mathbf{9}_{-1}+\mathbf{16}_{-1}+%
\mathbf{1}_{+3}+\mathbf{1}_{-3}  \notag \\
&&  \notag \\
&\rightarrow &\mathbf{1}_{+1}+\mathbf{1}_{+1}+\mathbf{1}_{+1}+\mathbf{8}%
_{v,+1}+\mathbf{8}_{c,+1}+\mathbf{8}_{s,+1}  \notag \\
&&+\mathbf{1}_{-1}+\mathbf{1}_{-1}+\mathbf{1}_{-1}+\mathbf{8}_{v,-1}+\mathbf{%
8}_{c,-1}+\mathbf{8}_{s,-1}+\mathbf{1}_{+3}+\mathbf{1}_{-3}.
\end{eqnarray}

\subsubsection{\label{Comments-Cartan-Subs}Comments on Cartan Subalgebras}

In the analysis made in Subsubsecs. \ref{First} and \ref{Second}, the $3$%
-dimensional non-compact Cartan subalgebra $\frak{H}_{3}$ (\ref
{Cartan-M-call}) of $\mathcal{M}$ (\ref{M-call}) is generated by a suitable
linear combination of the six $SO\left( 8\right) $-singlets (\ref
{six-SO(8)-singlets}). Thus, $\frak{H}_{3}$ is not the Lie algebra of the
group factor
\begin{equation}
\widetilde{H}_{3}:=\left[ SO\left( 1,1\right) \right] ^{3}\subsetneq
E_{6\left( -26\right) }\times SO\left( 1,1\right)  \label{H3-tilde}
\end{equation}
commuting with $SO\left( 8\right) $ in the branching (\ref{2-bis}), because
by definition for the Lie group $H_{3}$ generated by $\frak{H}_{3}$ it holds
that (recall definition (\ref{M-call-tilde})):
\begin{equation}
H_{3}\subsetneq \frac{E_{7\left( -25\right) }}{E_{6\left( -26\right) }\times
SO\left( 1,1\right) }=:\widetilde{\mathcal{M}},
\end{equation}
and, by definition:
\begin{equation}
\frak{T}\cap \frak{P}=0.
\end{equation}
As stated in the previous treatment, $\frak{H}_{3}$ can be extended to a $7$%
-dimensional maximal Cartan subalgebra $\frak{H}$ of $\frak{e}_{7\left(
-25\right) }$ by adding a $4$-dimensional maximal Cartan subalgebra of $%
\frak{so}\left( 8\right) $, which is clearly compact:
\begin{equation}
\frak{H}_{7}:=\frak{H}_{3}\oplus \frak{H}_{4},
\end{equation}
with signature $\left( +^{3},-^{4}\right) $ (indeed, as mentioned above,
compact generators are conventionally chosen with negative signature).

On the other hand, the factor $\widetilde{H}_{3}\equiv \left[ SO\left(
1,1\right) \right] ^{3}$ in (\ref{H3-tilde}) which commutes with $SO\left(
8\right) $ in the branching (\ref{2-bis}) can be extended to $\left[
SL\left( 2,\mathbb{R}\right) \right] ^{3}$, which with further branchings
gives rise to the (not maximal nor symmetric) embedding:
\begin{equation}
E_{6\left( -26\right) }\supsetneq \left[ SU\left( 2\right) \right]
^{4}\times \left[ SL\left( 2,\mathbb{R}\right) \right] ^{3},
\end{equation}
recently considered in \cite{ICL-4} within the quantum-informational
interpretation of $\mathcal{N}=2$, $D=4$ exceptional magic supergravity. As
done above for $\frak{H}_{3}$, the Lie algebra $\widetilde{\frak{H}}_{3}$ of
$\widetilde{H}_{3}$ can be extended to another $7$-dimensional maximal
Cartan subalgebra $\widetilde{\frak{H}}_{7}$ of $\frak{e}_{7\left(
-25\right) }$ by adding a $4$-dimensional maximal Cartan subalgebra of $%
\frak{so}\left( 8\right) $, which is clearly compact:
\begin{equation}
\widetilde{\frak{H}}_{7}:=\widetilde{\frak{H}}_{3}\oplus \frak{H}_{4},
\end{equation}
once again with signature $\left( +^{3},-^{4}\right) $.

\section{\label{Generalizations}Generalizations to groups of type $\mathbf{E_7}$}

The results derived until now hold \textit{at least} for the \textit{%
conformal non-compact} real forms of (\textit{non-degenerate} \cite
{Garibaldi}) \textit{simple} groups \textit{``of type }$E_{7}$\textit{''}
\cite{Brown,Duff-FD-1,FMY-FD-1,FMY-T-CV}. The first axiomatic
characterization of groups \textit{``of type }$E_{7}$\textit{''} through a
module (irreducible representation) was given in 1967 by Brown \cite{Brown}.
A group $G$ \textit{``of type }$E_{7}$\textit{''} is a Lie group endowed
with a representation $\mathbf{R}$ such that:

\begin{itemize}
\item  $\mathbf{R}$ is \textit{symplectic}, \textit{i.e.} (the subscripts ``$%
s$''\ and ``$a$''\ stand for symmetric and skew-symmetric throughout):
\begin{equation}
\exists !\mathbb{C}_{\left[ MN\right] }\equiv \mathbf{1\in R\times }_{a}%
\mathbf{R;}  \label{sympl-metric}
\end{equation}
$\mathbb{C}_{\left[ MN\right] }$ defines a non-degenerate skew-symmetric
bilinear form (\textit{symplectic product}); given two different charge
vectors $\mathcal{Q}_{x}$ and $\mathcal{Q}_{y}$ in $\mathbf{R}$, such a
bilinear form is defined as:
\begin{equation}
\left\langle \mathcal{Q}_{x},\mathcal{Q}_{y}\right\rangle \equiv \mathcal{Q}%
_{x}^{M}\mathcal{Q}_{y}^{N}\mathbb{C}_{MN}=-\left\langle \mathcal{Q}_{y},%
\mathcal{Q}_{x}\right\rangle .  \label{WW}
\end{equation}

\item  $\mathbf{R}$ admits a unique rank-$4$ completely symmetric primitive $%
G$-invariant structure, usually named $K$-tensor:
\begin{equation}
\exists !\mathbb{K}_{\left( MNPQ\right) }\equiv \mathbf{1\in }\left[ \mathbf{%
R\times R\times R\times R}\right] _{s}\mathbf{;}
\end{equation}
thus, by contracting the $K$-tensor with the same charge vector $\mathcal{Q}$
in $\mathbf{R}$, one can construct a rank-$4$ homogeneous $G$-invariant
polynomial (whose $\varsigma $ is the normalization constant):
\begin{equation}
\mathbf{q}\left( \mathcal{Q}\right) \equiv \varsigma \mathbb{K}_{MNPQ}%
\mathcal{Q}^{M}\mathcal{Q}^{N}\mathcal{Q}^{P}\mathcal{Q}^{Q},  \label{I4}
\end{equation}
which corresponds to the evaluation of the rank-$4$ symmetric invariant $%
\mathbf{q}$-structure induced by the $K$-tensor on four identical modules $%
\mathbf{R}$:
\begin{equation}
\mathbf{q}\left( Q\right) \equiv \left. \mathbf{q}\left( \mathcal{Q}_{x},%
\mathcal{Q}_{y},\mathcal{Q}_{z},\mathcal{Q}_{w}\right) \right| _{\mathcal{Q}%
_{x}=\mathcal{Q}_{y}=\mathcal{Q}_{z}=\mathcal{Q}_{w}\equiv \mathcal{Q}%
}\equiv \varsigma \left[ \mathbb{K}_{MNPQ}\mathcal{Q}_{x}^{M}\mathcal{Q}%
_{y}^{N}\mathcal{Q}_{z}^{P}\mathcal{Q}_{w}^{Q}\right] _{\mathcal{Q}_{x}=%
\mathcal{Q}_{y}=\mathcal{Q}_{z}=\mathcal{Q}_{w}\equiv \mathcal{Q}}.
\end{equation}
A famous example of \textit{quartic} invariant in $G=E_{7}$ is the \textit{%
Cartan-Cremmer-Julia} invariant\footnote{%
As also mentioned in \cite{Helenius}, it should be noted that the quartic
form is given incorrectly by Cartan; the error seems to have been first
observed by Freudenthal \cite{freudenthal-2}.} (\cite{Cartan}, p. 274),
constructed out of the fundamental representation $\mathbf{R}=\mathbf{56}$.

\item  If a trilinear map $T\mathbf{:R\times R\times R}\rightarrow \mathbf{R}
$ is defined such that:
\begin{equation}
\left\langle T\left( \mathcal{Q}_{x},\mathcal{Q}_{y},\mathcal{Q}_{z}\right) ,%
\mathcal{Q}_{w}\right\rangle =\mathbf{q}\left( \mathcal{Q}_{x},\mathcal{Q}%
_{y},\mathcal{Q}_{z},\mathcal{Q}_{w}\right) ,
\end{equation}
then it holds that:
\begin{equation}
\left\langle T\left( \mathcal{Q}_{x},\mathcal{Q}_{x},\mathcal{Q}_{y}\right)
,T\left( \mathcal{Q}_{y},\mathcal{Q}_{y},\mathcal{Q}_{y}\right)
\right\rangle = -2 \left\langle \mathcal{Q}_{x},\mathcal{Q}_{y}\right\rangle
\mathbf{q}\left( \mathcal{Q}_{x},\mathcal{Q}_{y},\mathcal{Q}_{y},\mathcal{Q}%
_{y}\right) .
\end{equation}
This last property makes the group of type $E_{7}$ amenable to a treatment
in terms of (rank-$3$) Jordan algebras and related Freudenthal triple
systems.
\end{itemize}

Remarkably, groups of type $E_{7}$, appearing in $D=4$ supergravity as $U$%
-duality groups, admit a $D=5$ uplift to groups of type $E_{6}$, as well as
a $D=3$ downlift to groups of type $E_{8}$. It should also be recalled that
split forms of exceptional Lie groups of type $E$ appear in the exceptional
Cremmer-Julia \cite{CJ-1} sequence $E_{11-D, \, \left( 11-D\right) }$ of $U$%
-duality groups of $M$-theory compactified on a $D$-dimensional torus, in $%
D=3,4,5$. Other sequences, composed by non-split, non-compact real forms of
exceptional groups, are also relevant to non-maximal supergravity in various
dimensions (see \textit{e.g.} the treatment in \cite{Exc-Reds}, also for a
list of related Refs.).

The connection of groups of type $E_{7}$ to supergravity can be summarized
by stating that all $2\leqslant \mathcal{N}\leqslant 8$-extended
supergravities in $D=4$ with symmetric scalar manifolds ${\frac{G_{4}}{H_{4}}%
}$ have $G_{4}$ of type $E_{7}$ \cite{Duff-FD-1,FMY-FD-1}. It is intriguing
to notice that the first paper on groups of type $E_{7}$ was written about a
decade before the discovery of of extended ($\mathcal{N}=2$) supergravity
\cite{FVN}, in which electromagnetic duality symmetry was observed \cite{FSZ}%
.

In particular, \textit{simple} $U$-duality groups of $\mathcal{N}=2$, $D=4$
theories with symmetric (vector multiplets') scalar manifolds (listed in
Table \ref{scalarmnf}) are \textit{conformal non-compact}, real forms of simple non-degenerate groups of type $E_{7}$, which are the conformal symmetry group of simple Euclidean Jordan algebras of rank 3 \cite{bart-sud}.

Furthermore, the results of Secs. \ref{sec:56} and \ref{manifest} also hold
for the relevant non-compact, real forms of (\textit{non-degenerate} \cite
{Garibaldi}) \textit{semi}-\textit{simple} groups \textit{of type }$E_{7}$
\cite{Brown,Duff-FD-1,FMY-FD-1,FMY-T-CV}, appearing in supergravity as
\textit{semi-simple} $U$- duality group of the infinite sequence of $%
\mathcal{N}=2$ theories, with scalar manifold given by:
\begin{equation}
\frac{SL\left( 2,\mathbb{R}\right) }{U\left( 1\right) }\times \frac{SO\left(
2,n\right) }{SO\left( 2\right) \times SO\left( n\right) },~n\geqslant 1,~%
\text{rank}=1+\text{min}\left( 2,n\right) ,  \label{N=2-infinite-seq}
\end{equation}
based on the \textit{semi-simple} rank-$3$ Jordan algebra $\mathbb{R}\oplus
\mathbf{\Gamma }_{1,n-1}$, where $\mathbf{\Gamma }_{1,n-1}$ stands for the
Jordan algebra of degree two with a quadratic form of Lorentzian signature $%
\left( 1,n-1\right) $, which is nothing but the Clifford algebra of $O\left(
1,n-1\right) $ \cite{JVNW}.
\begin{table}[t]
\begin{center}
\begin{tabular}{|c||c|c|c|}
\hline
$
\begin{array}{c}
\\
\frak{J}_{3}
\end{array}
$ & $
\begin{array}{c}
\\
G_{4}/H_{4} \\
~~
\end{array}
$ & $
\begin{array}{c}
\\
\mathbf{R} \\
~~
\end{array}
$ & $
\begin{array}{c}
\\
q \\
~~
\end{array}
$ \\ \hline\hline
$
\begin{array}{c}
\\
\frak{J}_{3}(\mathbb{O}) \\
~
\end{array}
$ & $\frac{E_{7\left( -25\right) }}{E_{6\left( -78\right) }\times U\left(
1\right) }~$ & $\mathbf{56}$ & $8~$ \\ \hline
$
\begin{array}{c}
\\
\frak{J}_{3}(\mathbb{H}) \\
~
\end{array}
$ & $\frac{SO^{\ast }\left( 12\right) }{SU\left( 6\right) \times U\left(
1\right) }$ & $\mathbf{32}$ & $4$ \\ \hline
$
\begin{array}{c}
\\
\frak{J}_{3}(\mathbb{C}) \\
~
\end{array}
$ & $\frac{SU\left( 3,3\right) }{SU\left( 3\right) \times SU\left( 3\right)
\times U(1)}$ & $\mathbf{20}$ & $2~$ \\ \hline
$
\begin{array}{c}
\\
\frak{J}_{3}(\mathbb{R}) \\
~
\end{array}
$ & $\frac{Sp\left( 6,\mathbb{R}\right) }{SU\left( 3\right) \times U\left(
1\right) }$ & $\mathbf{14}^{\prime }$ & $1$ \\ \hline
$
\begin{array}{c}
\\
\mathbb{R} \\
(T^{3}\text{~model})~~
\end{array}
$ & $\frac{SL\left( 2,\mathbb{R}\right) }{U\left( 1\right) }$ & $\mathbf{4}$
& $-2/3$ \\ \hline
\end{tabular}
\end{center}
\caption{Vector multiplets' \textit{symmetric} scalar manifolds (\ref
{M-gen-struct}) (up to possible finite factors in the stabiliser) of $%
\mathcal{N}=2$, $D=4$ supergravity models with \textit{simple} $U$-duality
groups (\textit{alias} \textit{conformal non-compact} real forms of \textit{%
non-degenerate} \protect\cite{Garibaldi}, \textit{simple} group \textit{of
type }$E_{7}$ \protect\cite{Brown,Duff-FD-1,FMY-FD-1,FMY-T-CV}), with
related \textit{simple} rank-$3$ Jordan algebra. The relevant symplectic
irrep. $\mathbf{R}$ of $G_{4}$ is also reported. $\mathbb{O}$, $\mathbb{H}$,
$\mathbb{C}$ and $\mathbb{R}$ respectively denote the four division algebras
of octonions, quaternions, complex and real numbers. Note that, with the
exception of the \textit{triality symmetric} $STU$ model \protect\cite{stu},
these models are all the ones for which the treatment of \protect\cite{Exc-Reds}
holds (see \textit{e.g.} Table 1 therein). The $D=5$ uplift of the $T^{3}$
model based on $\frak{J}_{3}=\mathbb{R}$ is the \textit{pure} $\mathcal{N}=2$, $D=5$
supergravity. $\frak{J}_3(\mathbb{H})$ is related to both $8$ and $24$
supersymmetries, because the corresponding supergravity theories are \textit{%
``twin''}, namely they share the very same bosonic sector \protect\cite{twin}%
. }
\label{scalarmnf}
\end{table}

In other words, at the group level, the results of Secs. \ref{sec:56} and
\ref{manifest} provide a manifestly $\left[ \text{mcs}\left( \text{Conf}%
\left( \frak{J}_{3}\right) \right) \right] $-covariant symplectic
frame for the \textit{special K\"{a}hler geometry} of the
corresponding symmetric, non-compact, vector multiplet's scalar
manifold (of Riemannian nature), whose coset structure reads (up to
possible finite factors in the stabilizer; see \textit{e.g.}
\cite{G-Lects,Small-Orbits-Phys}; for a comprehensive list of manifolds, see \textit{e.g.} \cite{LA08-Proc}):
\begin{equation}
\mathcal{M}_{\mathcal{N}=2}=\frac{\text{Conf}\left( \frak{J}_{3}\right) }{%
\text{mcs}\left( \text{Conf}\left( \frak{J}_{3}\right) \right) }.
\label{M-gen-struct}
\end{equation}
Here Conf$\left( \frak{J}_{3}\right) =$Aut$\left( \frak{M}\left( \frak{J}%
_{3}\right) \right) $ stands for the \textit{conformal group} of $\frak{J}%
_{3}$, which is nothing but the automorphism group of the \textit{%
Freudenthal triple system} $\frak{M}$ \cite{freudenthal} constructed on $%
\frak{J}_{3}$ itself. The relevant non-compact, real forms of mcs$\left[ \text{Conf}\left(\frak{J}_{3}\right) \right]/U(1)$ 
(namely, the U-duality symmetries in $D=5$) are the reduced structure algebras of
the corresponding ($q$-parametrized) simple, rank-3 Euclidean Jordan
algebras.

Up to symplectic re-parametrization, for the infinite sequence (\ref
{N=2-infinite-seq}) of $\mathcal{N}=2$ theories with semi-simple
$U$-duality group Conf$\left( \frak{J}_{3}\right) =SL\left(
2,\mathbb{R}\right) \times SO\left( 2,n\right) $, the results of
Secs. \ref{sec:56} and \ref{manifest}\ match the so-called
\textit{Calabi-Vesentini} $\mathcal{N}=2$ symplectic frame
\cite{CV,CDFVP} (see also \cite{FMY-T-CV} for a recent study), whose
(compact) manifest covariance is the maximal one:
\begin{equation}
\text{mcs}\left( \text{Conf}\left( \frak{J}_{3}\right) \right) =\text{mcs}%
\left( SL\left( 2,\mathbb{R}\right) \times SO\left( 2,n\right) \right)
=U\left( 1\right) \times SO\left( 2\right) \times SO\left( n\right) .
\end{equation}

All the vector multiplets' scalar manifolds of the aforementioned $\mathcal{N%
}=2$, $D=4$ supergravity theories related to cubic Euclidean Jordan algebras
are special K\"{a}hler, maximal, non-compact, symmetric cosets with
structure (\ref{M-gen-struct}), and have \textit{rank}\footnote{%
The \textit{rank} of a manifold is defined as the maximal dimension (in $%
\mathbb{R}$) of a Riemann-flat, totally geodesic sub-manifold of the
manifold itself (see \textit{e.g.} \cite{Helgason}, p. 209).} $3$ (except
the rank-$1$ case of $T^{3}$ model). They also are Einstein spaces, with
constant (negative) Ricci scalar curvature $R$:
\begin{equation}
R_{i\overline{j}}=\lambda g_{i\overline{j}}\Rightarrow R=\lambda n_{V},
\end{equation}
where $R_{i\overline{j}}$ is the special K\"{a}hler Ricci tensor, and the
real parameter $\lambda $ has been computed in \cite{CVP} (see also \cite
{Raju-1}):
\begin{equation}
\lambda =\left\{
\begin{array}{l}
-\frac{2}{3}n_{V}~\text{for:~}T^{3}~\text{model~}\left( n_{V}=1\right) \text{%
,~}STU~\text{model}~\left( n_{V}=3\right) \text{, and }\frak{J}_{3}(\mathbb{A})\text{-models~}\left( n_{V}=3q+3\right) ; \\
\\
-\frac{\left( n_{V}^{2}-2n_{V}+3\right) }{n_{V}}~\text{for~}\mathbb{R}\oplus
\mathbf{\Gamma }_{1,n-1}~\text{models~}\left( n_{V}=n+1\geqslant 2\right) .
\end{array}
\right.
\end{equation}

Similarly, also the results about the Iwasawa decomposition
worked out in Sec. \ref{Iwa-Exc} can be generalized \textit{at least} to the \textit{conformal
non-compact} real forms of (\textit{non-degenerate} \cite{Garibaldi})
\textit{simple} groups \textit{of type }$E_{7}$ \cite
{Brown,Duff-FD-1,FMY-FD-1,FMY-T-CV}, listed as $D=4$ $U$-duality groups $%
G_{4}$'s in Table \ref{scalarmnf}.

Indeed, in light of (\ref{deff-1})-(\ref{deff-4}), the chain of maximal and symmetric embeddings (\ref{second}) enjoys the following generalization:
\begin{eqnarray}
\text{Conf}\left( \frak{J}_{3}\left( \mathbb{A}\right) \right) &\supset &%
\text{Str}_{0}\left( \frak{J}_{3}\left( \mathbb{A}\right) \right) \times
SO\left( 1,1\right) ^{\prime }  \notag \\
&\supset &\text{Aut}\left( \frak{J}_{3}\left( \mathbb{A}\right) \right)
\times SO\left( 1,1\right) ^{\prime }  \notag \\
&\supset &SO\left( q+1\right) \times \mathcal{A}_{q}\times SO\left(
1,1\right) ^{\prime }  \notag \\
&\supset &SO\left( q\right) \times \mathcal{A}_{q}\times SO\left( 1,1\right)
^{\prime },  \label{second-generalized}
\end{eqnarray}
by recalling the definition introduced just below Eq. (\ref{dd-norm}), $q:=$%
dim$_{\mathbb{R}}\mathbb{A}=1$, $2$, $4$ and $8$ for $\mathbb{A}=\mathbb{R}$%
, $\mathbb{C}$, $\mathbb{H}$ and $\mathbb{O}$, respectively. We report the
symmetry groups of \textit{simple} rank-$3$ Euclidean Jordan algebras in
Table \ref{tabgro}. \textit{Mutatis mutandis} (also with the help of the Tables), the
treatment and the results of the whole Sec. \ref{Iwa-Exc} can be extended to
all $\mathcal{N}=2$, $D=4$ \textit{symmetric} supergravities reported in
Table \ref{scalarmnf} (but the $T^{3}$\ model).

\begin{table}[tbp]
\begin{equation*}
\begin{array}{|c||c|c|c|c|}
\hline
\frak{J}_{3}\left( \mathbb{A}\right) & $Aut$\left( \frak{J}_{3}\left(
\mathbb{A}\right) \right)=$mcs$(G_5) & $Str$_{0}\left( \frak{J}_{3}\left(
\mathbb{A}\right) \right)=G_5 & $Conf$\left( \frak{J}_{3}\left( \mathbb{A}%
\right) \right)=G_4 & $QConf$\left( \frak{J}_{3}\left( \mathbb{A}\right)
\right)=G_3 \\ \hline\hline
\mathbb{R} & Id & Id & Sl(2,\mathbb{R}) & G_{2(2)} \\ \hline
\frak{J}_{3}^{\mathbb{R}} & SO(3) & SL(3,\mathbb{R}) & Sp(6,\mathbb{R}) &
F_{4(4)} \\ \hline
\frak{J}_{3}^{\mathbb{C}} & SU(3) & SL(3,\mathbb{C}) & SU(3,3) & E_{6(2)} \\
\hline
\frak{J}_{3}^{\mathbb{H}} & USp(6) & SU^*(6) & SO^*(12) & E_{7(-5)} \\ \hline
\frak{J}_{3}^{\mathbb{O}} & F_{4(-52)} & E_{6(-26)} & E_{7(-25)} & E_{8(-24)}
\\ \hline
\end{array}
\end{equation*}
\caption{Invariance groups associated to \textit{simple} rank-$3$ Euclidean
Jordan algebras $\frak{J}_{3}\left( \mathbb{A}\right) $. Conf$\left( \frak{J}%
_{3}\left( \mathbb{A}\right) \right) $'s are \textit{conformal non-compact}
real forms of (\textit{non-degenerate} \protect\cite{Garibaldi}) \textit{%
simple} group \textit{of type }$E_{7}$ \protect\cite
{Brown,Duff-FD-1,FMY-FD-1,FMY-T-CV}. $G_{5}$, $G_{4}$ and $G_{3}$
respectively denote the $U$-duality groups of the corresponding supergravity
theories with $8$ supersymmetries in $D=5$, $4$ and $3$. The lower $4\times
4 $ part is known as the \textit{``Magic Square''}, due to its symmetry
along the diagonal (see \textit{e.g.} \protect\cite{GST}). }
\label{tabgro}
\end{table}

\begin{table}[h!]
\begin{center}
\begin{tabular}{|c|c|}
\hline
$
\begin{array}{c}
~q
\end{array}
$ & $\mathcal{A}_{q}$ \\ \hline\hline
$8~$ & \multicolumn{1}{|l|}{$-$} \\ \hline
$4$ & \multicolumn{1}{|l|}{$SO\left( 3\right) $} \\ \hline
$2$ & $SO\left( 2\right) $ \\ \hline
$1$ & $-$ \\ \hline
\end{tabular}
\end{center}
\caption{The extra commuting group $\mathcal{A}_{q}$ (see \textit{e.g.}
\protect\cite{CFMZ1-D=5}).}
\label{Aq}
\end{table}

\begin{table}[h!]
\begin{center}
\begin{tabular}{|c||c|c|c|}
\hline
$q$ & $\frak{tri}\left( \mathbb{A}\right) ~$ & $SO\left( q\right) \times
\mathcal{A}_{q}$ & Aut$\left( \mathbf{t}\left( \mathbb{A}\right) \right) $
\\ \hline
$
\begin{array}{c}
\\
1
\end{array}
$ & $\left\{ 0\right\}$ & $Id$ & $\left\{ \left( g_{1},g_{2},g_{3}\right)
\in \left[ O\left( 1\right) \right] ^{3}:g_{1}g_{2}g_{3}=1\right\}~$ \\
\hline
$
\begin{array}{c}
\\
2 \\
~
\end{array}
$ & $\left[ \frak{u}\left( 1\right) \right] ^{2}~$ & $SO\left( 2\right)
\times SO\left( 2\right) \sim \left[ U\left( 1\right) \right] ^{2}~$ & $%
\left\{ \left( g_{1},g_{2},g_{3}\right) \in \left[ U\left( 1\right) \right]
^{3}:g_{1}g_{2}g_{3}=1\right\} \times \mathbb{Z}_{2}$ \\ \hline
$
\begin{array}{c}
\\
4 \\
~
\end{array}
$ & $\left[ \frak{usp}\left( 2\right) \right] ^{3}$ & $SO\left( 4\right)
\times SO\left( 3\right) \sim \left[ SU\left( 2\right) \right] ^{3}$ & $%
\left[ USp\left( 2\right) \right] ^{3}/\left\{ \pm \left( 1,1,1\right)
\right\}~$ \\ \hline
$
\begin{array}{c}
\\
8 \\
~
\end{array}
$ & $\frak{so}\left( 8\right)~$ & $SO\left( 8\right)~$ & $Spin\left(
8\right)~$ \\ \hline
\end{tabular}
\end{center}
\caption{The Lie algebra $\frak{tri}\left( \mathbb{A}\right) $ of the
automorphism group Aut$\left( \mathbf{t}\left( \mathbb{A}\right) \right) $
of the \textit{normed triality} $\mathbf{t}\left( \mathbb{A}\right) $ over
the division algebra $\mathbb{A}$, and the group $SO\left( q\right) \times
\mathcal{A}_{q}$, in terms of the parameter $q$. See \textit{e.g.}
\protect\cite{Baez}, in particular Eqs. (5) and (21) therein.}
\label{triA}
\end{table}

\begin{table}[tbp]
\begin{center}
\begin{tabular}{|c||c|}
\hline
$
\begin{array}{c}
\\
A
\end{array}
$ & $
\begin{array}{c}
\\
\mathcal{O}_{p=2,BPS}=\frac{\frak{J}_{3}\left( \mathbb{A}\right) }{\mathcal{G%
}_{p=2}\left( \frak{J}_{3}\left( \mathbb{A}\right) \right) } \\
~~
\end{array}
$ \\ \hline
$
\begin{array}{c}
\\
\mathbb{O} \\
~
\end{array}
$ & $\frac{E_{7\left( -25\right) }}{SO\left( 8\right) }$ \\ \hline
$
\begin{array}{c}
\\
\mathbb{H} \\
~
\end{array}
$ & $\frac{SO^{\ast }(12)}{\left[ SU\left( 2\right) \right] ^{3}}$ \\ \hline
$
\begin{array}{c}
\\
\mathbb{C} \\
~
\end{array}
$ & $\frac{SU\left( 3,3\right) }{\left[ U(1)\right] ^{2}}$ \\ \hline
$
\begin{array}{c}
\\
\mathbb{R} \\
~
\end{array}
$ & $Sp\left( 6,\mathbb{R}\right) $ \\ \hline
\end{tabular}
\end{center}
\caption{BPS generic charge orbits of $2$-centered extremal black holes in $%
\mathcal{N}=2$, $d=4$ magical models. Conf$\left( \frak{J}_{3}\left( \mathbb{%
A}\right) \right) $ denotes the ``conformal'' group of $\frak{J}_{3}\left(
\mathbb{A}\right) $ (see \textit{e.g.} \protect\cite{G-Lects}) \protect\cite{Irred-1}.}
\label{BPS}
\end{table}

The extra factor group $\mathcal{A}_{q}$, which exists only for $q=2$ and $%
q=4$, is reported in Table \ref{Aq}; in \cite{CFMZ1-D=5}, its appearance was
observed within the study of the charge orbits of asymptotically flat $0$-
(black holes) and $1$- (black strings) branes in minimal magical
Maxwell-Einstein supergravity theories in $D=5$ space-time dimensions. We
note that $\mathcal{A}_{q}$ is related to $\widehat{G}_{cent}$ and $%
G_{paint} $ (Lie groups usually introduced in the treatment of \textit{%
supergravity billiards} and timelike Kaluza-Klein reductions; for \ recent
treatment and set of related Refs., see \textit{e.g.} \cite{Trig-Berg}; see
also Table 5 therein, also for subtleties concerning the case $q=8$ in $%
D=5,6 $) as follows \cite{CFMZ1-D=5}:
\begin{eqnarray}
D &=&5,6:\widehat{G}_{cent}=SO\left( 1,1\right) \times SO\left( q-1\right)
\times \mathcal{A}_{q}; \\
D &=&3,4:\widehat{G}_{cent}=G_{paint}=SO\left( q\right) \times \mathcal{A}%
_{q}.  \label{D=3,4}
\end{eqnarray}
According to \cite{Armenians-1}, $\mathcal{A}_{q}$ can be related to the
structure of the Hopf maps, chiral Weyl spinors and division algebras; we
hope to study this intriguing connection in future investigations, also
along the lines of \cite{Baez-Huerta}.

Extending the considerations made above on $SO\left( 8\right) $, it can be
observed that $SO\left( q\right) \times \mathcal{A}_{q}$ shares the same Lie
algebra $\frak{tri}\left( \mathbb{A}\right) $ of Aut$\left( \mathbf{t}\left(
\mathbb{A}\right) \right) $, which is the automorphism group of the \textit{%
normed triality} over the division algebra $\mathbb{A}$ (see \textit{e.g.}
\cite{Baez}); see Table \ref{triA}.

Besides this fact, it is intriguing to notice that $SO\left( q\right) \times
\mathcal{A}_{q}$ appears in \textit{at least} three (apparently unrelated)
contexts:

\begin{enumerate}
\item  As $\widehat{G}_{cent}=G_{paint}$ in $D=3,4$, as given by (\ref{D=3,4}%
) (see \textit{e.g.} \cite{CFMZ1-D=5,Trig-Berg}).

\item  As stabilizer group $\mathcal{G}_{p=2}\left( \frak{J}_{3}\left(
\mathbb{A}\right) \right) $ of BPS generic charge orbits of $2$-centered
extremal black holes in $\mathcal{N}=2$, $D=4$ magical models, as derived in
\cite{Irred-1}, and reported in Table \ref{BPS}.

\item  According to (\ref{second-generalized}), as group of maximal
manifest covariance of the Iwasawa decomposition of the (vector multiplets')
scalar manifold of $\mathcal{N}=2$, $D=4$ magical models, whose $U$-duality
groups are (some instances of) \textit{conformal non-compact} real forms of (%
\textit{non-degenerate} \cite{Garibaldi}) \textit{simple} groups \textit{of
type }$E_{7}$ \cite{Brown,Duff-FD-1,FMY-FD-1,FMY-T-CV}.\medskip
\end{enumerate}

\section{\label{Conclusion}Conclusion}

The present investigation, and in particular the generalizations discussed
in Sec. \ref{Generalizations}, pave the way to
a number of interesting further developments. We list a selection of them
below.

Starting with the treatment given in Sec. \ref{manifest}, it should be
pointed out that a more explicit expression of the symplectic frame
determined by the comparison of (\ref{dcoset})-(\ref{dcoset2}) with the
general formul\ae\ (\ref{M-call-par})-(\ref{K}) of special K\"{a}hler
geometry would be needed also in order to check that the prepotential $F$
does \textit{not} exist in the symplectic frame introduced in Secs. \ref
{sec:56} and \ref{manifest}, which can be considered the \textit{analogue of
the Calabi-Vesentini} one \cite{CV,CDFVP} for \textit{non-degenerate},
\textit{conformal non-compact}, \textit{simple} groups of type $E_{7}$.

Furthermore, it would be interesting to extend the maximally
manifestly-covariant symplectic frame and/or the Iwasawa symplectic frame,
respectively introduced in Secs. \ref{sec:56}-\ref{manifest} and in Sec. \ref
{Iwa-Exc}, to

\begin{itemize}
\item  \textit{compact} groups of type $E_{7}$;

\item  other \textit{non-compact} real forms of groups of type $E_{7}$
(possibly related to $\mathcal{N}>2$-extended supergravity theories), also
in relation to rank-$3$ Jordan algebras on \textit{split} forms of division
algebras (for a recent treatment of the non-supersymmetric cases of $\frak{J}%
_{3}(\mathbb{H}_{s})$ and $\frak{J}_3(\mathbb{C}_{s})$, see \textit{e.g.}
\cite{Small-Orbits-Maths});

\item  other classes of supergravities, such as $\mathcal{N}=2$, $D=4$ with
homogeneous \textit{non-symmetric} scalar manifolds \cite{dWVP-cubic,dWVVP}.
\end{itemize}

Furthermore, considering the generalizations of the Iwasawa parametrization
discussed in Sec.~\ref{Generalizations}, it would be interesting to
explore its extension also to theories related to \textit{semi-simple} rank-$%
3$ Jordan algebras, such as $\mathbb{R}\oplus \mathbf{\Gamma }_{m,n}$ (for $%
m=1$, recall (\ref{N=2-infinite-seq})), where $\mathbf{\Gamma }_{m,n}$
stands for the rank-$2$ Jordan algebra with a quadratic form of Lorentzian
signature $\left( m,n\right) $, which is nothing but the Clifford algebra of
$O\left( m,n\right) $ \cite{JVNW}.

Concerning the generalization to $\mathcal{N}>2$-extended
supergravities, it would be interesting to compare the application
of the Iwasawa decomposition under consideration to the case of
$\mathbb{O}_{S}$ with the Iwasawa parametrization of
$\mathcal{M}_{\mathcal{N}=8}$ (\ref {M-call-N=8}) studied in
\cite{Iwa-N=8}, whose manifest maximal
(non-compact) covariance is $SL\left( 7,\mathbb{R}\right) $, with mcs $%
SO\left( 7\right) $. In this respect, the following remark made in \cite
{Irred-1} should be relevant : as it holds for the stabilizer of $\mathcal{O}%
_{\mathcal{N}=2,\frak{J}_{3}\left( \mathbb{O}\right) ,\text{BPS},p=2}$ (see
Table \ref{BPS}), the Lie algebra $\frak{so}\left( 8\right) $ of the stabilizer of
the $2$-centered orbit \cite{Irred-1}
\begin{equation}
\mathcal{O}_{\mathcal{N}=2,J_{3}^{\mathbb{O}},\text{nBPS},p=2,\mathbf{I}}=%
\frac{E_{7\left( -25\right) }}{SO\left( 8\right) }
\end{equation}
is nothing but the Lie algebra $\frak{tri}\left( \mathbb{O}\right) $ of the
automorphism group Aut$\left( \mathbf{t}\left( \mathbb{O}\right) \right) $
of the \textit{normed triality} over $\mathbb{O}$ (see Table \ref{triA}). It is here
worth observing that the Lie algebra $\frak{so}\left( 4,4\right) $ of the
stabilizer of the $2$-centered orbit \cite{Irred-1}
\begin{equation}
\mathcal{O}_{\mathcal{N}=8,\frac{1}{8}\text{-BPS},p=2,\mathbf{I}}=\frac{%
E_{7\left( 7\right) }}{SO\left( 4,4\right) }
\end{equation}
enjoys an analogous interpretation as the Lie algebra $\frak{tri}\left(
\mathbb{O}_{s}\right) $ of the automorphism group Aut$\left( \mathbf{t}%
\left( \mathbb{O}_{S}\right) \right) $ of the \textit{normed triality} over $%
\mathbb{O}_{S}$. On the other hand, a similar interpretation does not seem to hold for the stabilizer of the $2$-centered orbit \cite{Irred-1}
\begin{equation}
\mathcal{O}_{\mathcal{N}=8,\frac{1}{8}\text{-BPS},p=2,\mathbf{II}}=\frac{%
E_{7\left( 7\right) }}{SO\left( 5,3\right) },
\end{equation}
as well as for the stabilizer of the $2$-centered orbit \cite{Irred-1}
\begin{equation}
\mathcal{O}_{\mathcal{N}=2,J_{3}^{\mathbb{O}},\text{nBPS},p=2,\mathbf{II}}=%
\frac{E_{7\left( -25\right) }}{SO\left( 7,1\right) }.  \label{jazz-1}
\end{equation}
However, it is intriguing to note that the maximal manifest compact
covariance $SO\left( 7\right) =$mcs$\left( SO\left( 7,1\right) \right) $%
\linebreak $=$mcs$\left( SL\left( 7,\mathbb{R}\right) \right) $ exhibited by
the Iwasawa parametrization of $\mathcal{M}_{\mathcal{N}=8}$ (\ref
{M-call-N=8}) \cite{Iwa-N=8} may provide a clue for the stabilizer of $%
\mathcal{O}_{\mathcal{N}=2,J_{3}^{\mathbb{O}},\text{nBPS},p=2,\mathbf{II}}$ (%
\ref{jazz-1}). We leave to future studies the in-depth investigation of
these fascinating connections, here just briefly outlined.

Finally, we would like to put forward an hint\footnote{%
We are grateful to Augusto Sagnotti for an enlightening remark on this.} for
a further physical application. It should be observed that the Iwasawa coset
decomposition can yield a nilpotent algebra exhibiting the same symmetry of
the systems recently discussed in \cite{FST} in the framework of
supergravity theories timelike-reduced down to $D=3^{\ast }$ dimensions.
Thus, it would be interesting to investigate the possible Lax pair
structures hidden in the Iwasawa formalism, which might allow for a more
explicit integration procedure within the $D=3^{\ast }$ nilpotent orbits
formalism of \cite{FST}. We hope to report on this intriguing connection in
future studies.

\section*{Acknowledgments}

We would like to thank Leron Borsten, Sergio Ferrara and Augusto Sagnotti
for enlightening discussions. We also thank Francesco Dalla Piazza for technical support.

The work of B.L.C. has been supported in part by the European Commission
under the FP7-PEOPLE-IRG-2008 Grant No. PIRG04-GA-2008-239412 \textit{%
``String Theory and Noncommutative Geometry''} (STRING).

\end{document}